 \documentclass[final,authoryear,5p,times,twocolumn]{elsarticle}
\pdfoutput=1

\usepackage{amssymb}
\usepackage{lscape}

\journal{Icarus}
\def\8p{8P/Tuttle}
\def\pdb{Plateau de Bure}

\def\tdeg{$^{\circ}$}

\begin{document}

\begin{frontmatter}



\title{Gas and dust productions of comet 103P/Hartley~2 from millimetre observations: interpreting rotation-induced time variations}

\author[iram,eso,ira]{J. Boissier}
\author[obsp]{D. Bockel\'ee-Morvan}
\author[obsp]{N. Biver}
\author[obsp]{P. Colom}
\author[obsp]{J. Crovisier}
\author[obsp]{R. Moreno}
\author[obsp,GS]{V. Zakharov}
\author[obsm]{O.Groussin}
\author[obsm]{L. Jorda}
\author[cti]{D.C. Lis}
\address[iram]{IRAM, 300 rue de la piscine, 38406 Saint Martin d'H\`eres, France (e-mail: boissier@iram.fr)}
\address[eso]{ESO, Karl Schwarzschild Str. 2, 85748 Garching bei Muenchen, Germany}
\address[ira]{Istituto di Radioastronomia - INAF, Via Gobetti 101, Bologna, Italy}

\address[obsp]{LESIA -- Observatoire de Paris, CNRS, UPMC, Universit´e Paris-Diderot, 5 place Jules Janssen, 92195 Meudon, France}
\address[GS]{Gordien Strato, S.A.R.L., 91370, Verrieres-le-Buisson, France}
\address[obsm]{Aix Marseille Universit\'e, CNRS, LAM (Laboratoire d'Astrophysique de Marseille) UMR 7326, 13388, Marseille, France}
\address[cti]{California Institute of Technology, Cahill Center for Astronomy and Astrophysics 301-17, Pasadena, CA 91125, USA}

\begin{abstract}
Comet 103P/Hartley 2  made a close approach to the Earth in October 2010. It was the target of an extensive observing campaign including ground- and orbit-based observatories and was visited by the Deep Impact spacecraft in the framework of its mission extension EPOXI.
We present observations of HCN and CH$_3$OH emission lines conducted with the IRAM Plateau de Bure interferometer on 22--23, 28 October and 4, 5 November 2010 at  1.1, 1.9 and 3.4~mm wavelengths. The thermal emission from the dust coma and nucleus is detected simultaneously. Interferometric images with unprecedented spatial resolution of $\sim$100 to $\sim$500 km are obtained.  
A sine-wave like variation of the thermal continuum is observed in the 23 October data, that we associate with the nucleus thermal light curve. The nucleus contributes up to 30--55 \% of the observed continuum emission.
The dust thermal emission is used to measure the dust production rate. The inferred large dust-to-gas ratio (in the range 2--6) can be explained by the unusual  activity of the comet for its size, which allows decimeter size particles and large boulders to be entrained by the gas due to the small nucleus gravity.
The rotational temperature of CH$_3$OH is measured with beam radii from $\sim$150~km to $\sim$1500~km. We attribute the increase from $\sim$35~K to $\sim$46~K with increasing beam size to radiative processes.
The HCN production rate displays strong rotation-induced temporal variations, varying  from  $\sim$0.3$\times$10$^{25}$~s$^{-1}$ to $\sim$2.0$\times$10$^{25}$~s$^{-1}$ in the 4--5 November period. 
The HCN production curve, as well as the CO$_2$ and H$_2$O production curves measured by EPOXI, are interpreted with a geometric model which takes into account the complex rotational state of 103P/Hartley 2 and its shape. The HCN and H$_2$O production curves are in phase, showing that these molecules have common sources. The $\sim$ 1.7 h delay, in average, of the HCN and H$_2$O production curves with respect to the CO$_2$ production curve suggests that HCN and H$_2$O  are mainly produced by subliming icy grains. The scale length of production of HCN is determined to be on the order of 500--1000 km, implying a mean velocity of 100--200 m s$^{-1}$ for the icy grains producing HCN. From the time evolution of the insolation of the nucleus, we show that the CO$_2$ production is modulated by the insolation of the small lobe of the nucleus. The three-cycle pattern of the production curves reported earlier is best explained by an overactivity of the small lobe in the longitude range 0--180$^{\circ}$. The good correlation
  between the insolation of the small lobe and CO$_2$ production is consistent with CO$_2$ being produced from small depths below the surface. The time evolution of the velocity offset of the HCN lines, as well as the displacement of the HCN photocenter in the interferometric maps, are overall consistent with this interpretation. Other localized sources of gas on the nucleus surface are also suggested. 

\end{abstract}

\begin{keyword}
Comet 103P/Hartley 2 \sep Comets, coma \sep Comets, dust \sep Radio observations
\end{keyword}

\end{frontmatter}

% \linenumbers

\section{Introduction}
\label{sec-intro}

Interferometric observations of comets at millimetric wavelengths have demonstrated their usefulness to study the inner coma. They are well suited to determine the origin of the molecules observed in the coma that can be released from the nucleus surface or by distributed sources \citep[e.g., from photodissociation of parent molecules or the sublimation of grains;][]{wri+98,mil+06,boi+07,boc+10}. Interferometric maps can also reveal the presence of jets or anisotropies in the outgassing pattern \citep{boi+07,boc+09hb,boi+10}. In addition, the continuum emission from the dust coma and the nucleus can be observed simultaneously with emission lines \citep{alt+99,boi+09emp,boi+11}.

The 2010 apparition of 103P/Hartley~2 was a rare opportunity to undertake sensitive observations of a Jupiter-family comet with interferometric techniques. This Jupiter-family comet made a close approach to the Earth on 21 October 2010 at a geocentric distance $\Delta$ = 0.12~AU. On 4 November 2010, it was visited by the Deep Impact spacecraft in the framework of the EPOXI mission \citep{ahe+11}.
In order to support the EPOXI mission, a worldwide observing campaign was organized, including ground- and space-based instruments in all  wavelength ranges \citep{mee+11}. We present here the observations  carried out in October and November 2010 with the  Plateau de Bure interferometer operated by the Institut de Radioastronomie Millim\'etrique (IRAM). We also consider observations performed with the IRAM 30-m telescope and the Caltech Submillimeter Observatory (CSO) \citep{biv+11dps,dra+12}.

Because of the close distance between 103P/Hartley 2 and the Earth, the continuum radiation from both the dust coma and the nucleus was detected in interferometric mode. The dust continuum emission, which probes the thermal radiation from millimeter-sized particles, was used to measure the dust production rate \citep[e.g.,][]{jewluu90,jewluu92}. 
  The contribution of the nucleus thermal radiation to the detected radiation could be properly evaluated, and its thermal lightcurve be investigated, thanks to the detailed characterization of the nucleus from EPOXI and radar investigations \citep{ahe+11,har+11rad}. 

We obtained interferometric maps of HCN lines, as well as single-dish data usefully complementing in time coverage the extensive data set of \citet{dra+12}. EPOXI observed rotation-induced temporal variations of the gaseous production, and revealed that most of the activity was taking place at the small end of the nucleus \citep{ahe+11}. We analyze the HCN production curves, Doppler line shifts, and interferometric maps using a model which takes into account the nucleus shape and its complex rotation state in order to identify the sources of HCN. % %

The emission lines of methanol in the millimeter-wave range are frequently used to derive  beam-averaged rotational temperatures in cometary atmospheres. We present here the analysis of the interferometric and single-dish observations of a group of E-type methanol  lines around 157.2~GHz to estimate the temperature at different distances from the nucleus of Hartley~2.

Radar observations of 103P/Hartley 2 showed the presence of large grains ($>$cm) in the coma \citep{har+11rad}. Large ice chunks were individually seen near the nucleus by EPOXI \citep{ahe+11}. We studied the brightness distribution of HCN lines to investigate whether grains are the main source of HCN in the coma. The anisotropy of the outgassing is taken into account in this study on the basis of simple simulations.

The observations are described in Sect.~\ref{sec-obs}. The continuum and spectral data are analyzed in Sect.~\ref{sec-res-cont} and Sect.~\ref{sec-resu-line}, respectively. In Sect.~\ref{sec-jet}, we propose an interpretation of the HCN temporal variations related to the complex rotation state of the nucleus. The radial distribution of the HCN molecules derived from the interferometric maps is studied in Sect.~\ref{sec-radext}. A summary follows in Sect.~\ref{sec-sum}.

\section{Observations}
\label{sec-obs}

\subsection{Description}

\begin{table*}
\begin{center}
\caption{Log of the Plateau de Bure observations of comet 103P/Hartley 2}
\label{tab-obs}
\begin{tabular}{l  c c c c c c c c}
\hline
\noalign{\smallskip}
Date & Line  &Freq.  & \multicolumn{3}{c}{Reference position and UT time$^a$} & $\Delta$ & $r_h$ & $\phi^b$  \\
UT & & GHz & RA (h:min:s) & Dec (\tdeg:$'$:$''$)& h & AU & AU & \\
\hline
\noalign{\smallskip}
\hline
\noalign{\smallskip}
22.904--23.321 Oct.\phantom{.} 2010           & HCN $J$(1--0)       & \phantom{1}88.6 & 05:58:08.02 & 32:52:17.8&  8.0  &  0.122 & 1.06 & 54$^{\circ}$  \\
27.946--28.379 Oct.\phantom{.} 2010           & CH$_3$OH $J_0-J_{-1}$& 157.2           & 06:35:47.06 & 20:38:57.6&  9.0  &  0.131 & 1.05 & 57$^{\circ}$  \\
\phantom{0}4.042--4.342 \phantom{0}Nov. 2010 & HCN $J$(1--0)        & \phantom{1}88.6 & 07:07:24.74 & 06:58:24.5&  8.0   &  0.154 & 1.06 & 59$^{\circ}$  \\
\phantom{0}4.992--5.333 \phantom{0}Nov. 2010  & HCN $J$(3--2)        & 265.9           & 07:10:38.36 & 05:22:20.6&  8.0   &  0.158 & 1.06 & 59$^{\circ}$  \\

\hline
\noalign{\smallskip}
\end{tabular}
\end{center}
$^a$ Reference position (0,0) of interferometric maps at the given UT time, in apparent geocentric coordinates, corresponding to the Horizons/JPL ephemeris which takes into account EPOXI astrometry (Giorgini et al., \citeyear{gior+12}).
$^b$ Phase angle.
\end{table*}

The IRAM interferometer  is an array of six 15-m antennas located at the \pdb{}, in the French Alps. 
Comet 103P/Hartley 2 was observed  at  wavelengths of 3.4, 1.9, and 1.1~mm (88.6, 157.2 and 265.9 GHz, respectively) with  this instrument between 23 October 2010 and 5 November 2010. 
The main parameters of the observations (geocentric and heliocentric distances, ephemeris, and phase angle) are reported in Table~\ref{tab-obs}.

The array was  in its compact configuration with baseline lengths ranging from  $\sim$20 to $\sim$100~m on 23 October 2010, and from $\sim$20 to $\sim$180~m on the other dates. 
The resulting synthesized beam diameters (Full width at half-maximum, FWHM) are  $\sim$1 to 5$''$ (see beam sizes in Table~\ref{tab-fluxes-line}), depending on the observing frequency, which corresponds to  projected distances at the comet from $\sim$100 to 500~km. 
The wideband correlator (WIDEX) was dedicated to continuum observations using a total bandwidth of 3.6~GHz in two orthogonal linear polarizations, whereas the narrowband correlator was set up to observe molecular emission lines in the two polarizations.

In addition to interferometric data, single-dish spectra were acquired by the six antennas throughout the observing periods at a time interval of $\sim$40~min, and with an integration time on source of 2 min. 
To cancel the sky contribution, the position switching mode (On-Off) was used 
with the Off position  at a distance of~5$'$ in azimuth from the comet. 
The forward and main beam efficiencies of the antennas are 0.95 and 0.85, respectively, at 88.6 GHz, 0.93 and 0.67, respectively, at 157.2 GHz, and 0.86 and 0.46, respectively, at 265.9 GHz. This results in conversion factors of 21, 26 and 35~Jy~K$^{-1}$ at 88.6, 157.2, and 265.9 GHz, respectively.
The flux calibration uncertainty of the On-Off measurements is on the order of 10\%.
The whole calibration process was performed using the GILDAS software packages developed by IRAM \citep{pet05}. The single-dish data obtained independently from the six antennas were averaged.

\paragraph{23 October 2010} Between 22 October 21 h UT to 23 October 8 h UT, i.e., two days after its closest approach to Earth, we observed the $J$(1--0) line of HCN at 88.6 GHz with a channel spacing of 40 kHz
 (corresponding to a spectral resolution of $\sim$0.2 km s$^{-1}$).
The weather conditions were excellent: a system temperature around 80~K and a phase rms on the order of 10\tdeg.
 The gain calibration sources (observed every 30 min to monitor the instrumental and atmospheric phase and amplitude variations) were 0552+398 and 0548+378. MWC349 was used to determine the absolute flux scale with a 1$\sigma$ uncertainty of $\sim$5\%.

\paragraph{28 October 2010} From 27 October 22~h UT to 28 October 9~h UT, we observed several  methanol rotational lines (E transitions $J$$_0$--$J$$_{-1}$, $J$=1--6) around 157.2~GHz. 
Several narrow correlator units were used with channel spacings of 39 and 156 kHz, corresponding to effective resolutions of  0.1 and 0.4~km~s$^{-1}$, respectively.
The interferometric data acquired from 27 October 23~h UT to 28 October 1.5~h UT under poor atmospheric phase conditions are not usable.
For the remaining part, from 2.5 to 9~h UT, the conditions were acceptable with a system temperature around 150~K and a phase rms of $\sim$40\tdeg.
The gain calibration source was 0657+172 and MWC349 was used to determine the absolute flux scale with an accuracy of $\sim$10\%. 

\paragraph{4 November 2010}
The HCN $J$(1--0) line was observed with a channel spacing of 40 kHz between 1 and 8.5~h~UT, i.e., a few hours before the EPOXI closest approach  on 4 November, 14~h UT \citep{ahe+11}.
The system temperature was about 90~K and the maximum phase rms was about 30\tdeg.
The gain calibration source was 0736+017 and MWC349 was used to determine the absolute flux scale with an uncertainty of $\sim$5\%. 

\paragraph{5 November 2010}
The HCN $J$(3--2) line at 265.9~GHz was observed with a channel spacing of 40 kHz (corresponding to an effective resolution of $\sim$0.06 km s$^{-1}$) between 0 and 8.5~h~UT, i.e., ten hours after the EPOXI closest approach. 
The weather conditions were poor: the system temperature was of the order of 500~K and the maximum phase rms was about 60\tdeg.
The gain calibration source was 0748+126 and 3C454.3 was used to determine the absolute flux scale with an uncertainty of $\sim$15\%. 

\subsection{On-Off spectra}
\label{sec-oo}
\begin{figure}
 \begin{center}
\includegraphics[angle=0,width=\columnwidth]{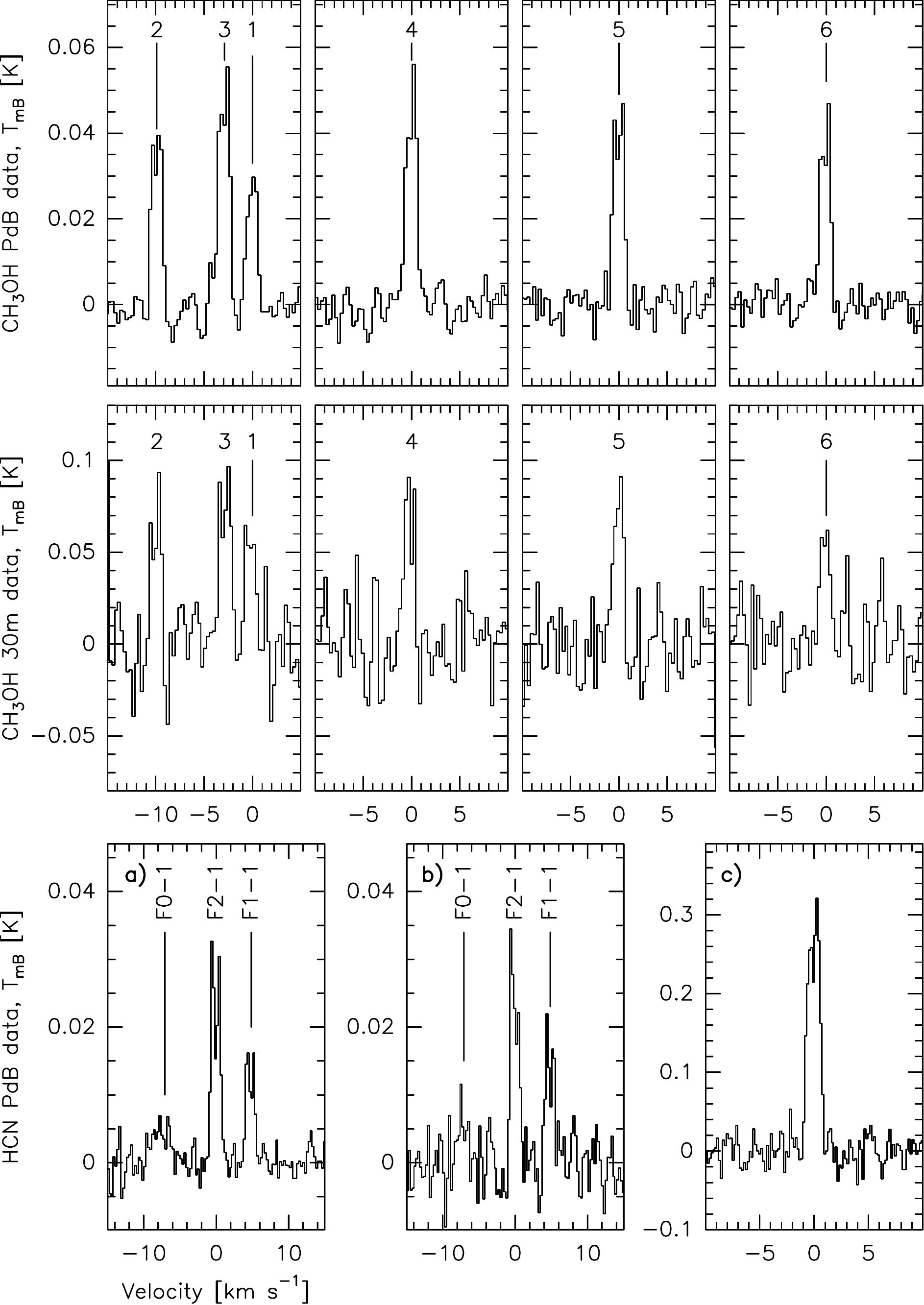} 
\caption{On-Off spectra observed in comet 103P/Hartley 2. 
{\bf Upper row:} methanol 157.2 GHz lines observed with the IRAM Plateau de Bure antennas on 28 October 2010. The total integration time is 32~min on source spread over a 9~h period.
{\bf Middle row:} methanol 157.2 GHz lines observed with the IRAM 30-m telescope on 28 October 2010. The total integration time is 19~min on source spread over a 1.15~h period. CH$_3$OH lines  are labelled from 1 to 6; the correspondence between labels and transitions is given in Table~\ref{tab-fluxes-line}. 
{\bf Lower row:} HCN spectra observed at the Plateau de Bure.
{\bf a)} HCN $J$(1--0) line at 88.6 GHz observed on 23 October 2010. The total integration time is 40 min spread over a 10~h interval. 
{\bf b)} HCN $J$(1--0) line at 88.6 GHz observed on 4 November 2010. The total integration time is 22 min spread over a 7~h interval. 
{\bf c)} HCN $J$(3--2) line at 265.9 GHz observed on 5 November 2010. The total integration time is 31 min spread over a 8~h interval. 
The hyperfine components of the HCN $J$(1--0) line are identified in frames {\bf a)} and {\bf b)}.}

\label{fig-oo}
 \end{center}
 \end{figure}

\begin{figure}
 \begin{center}
 \includegraphics[width=8cm]{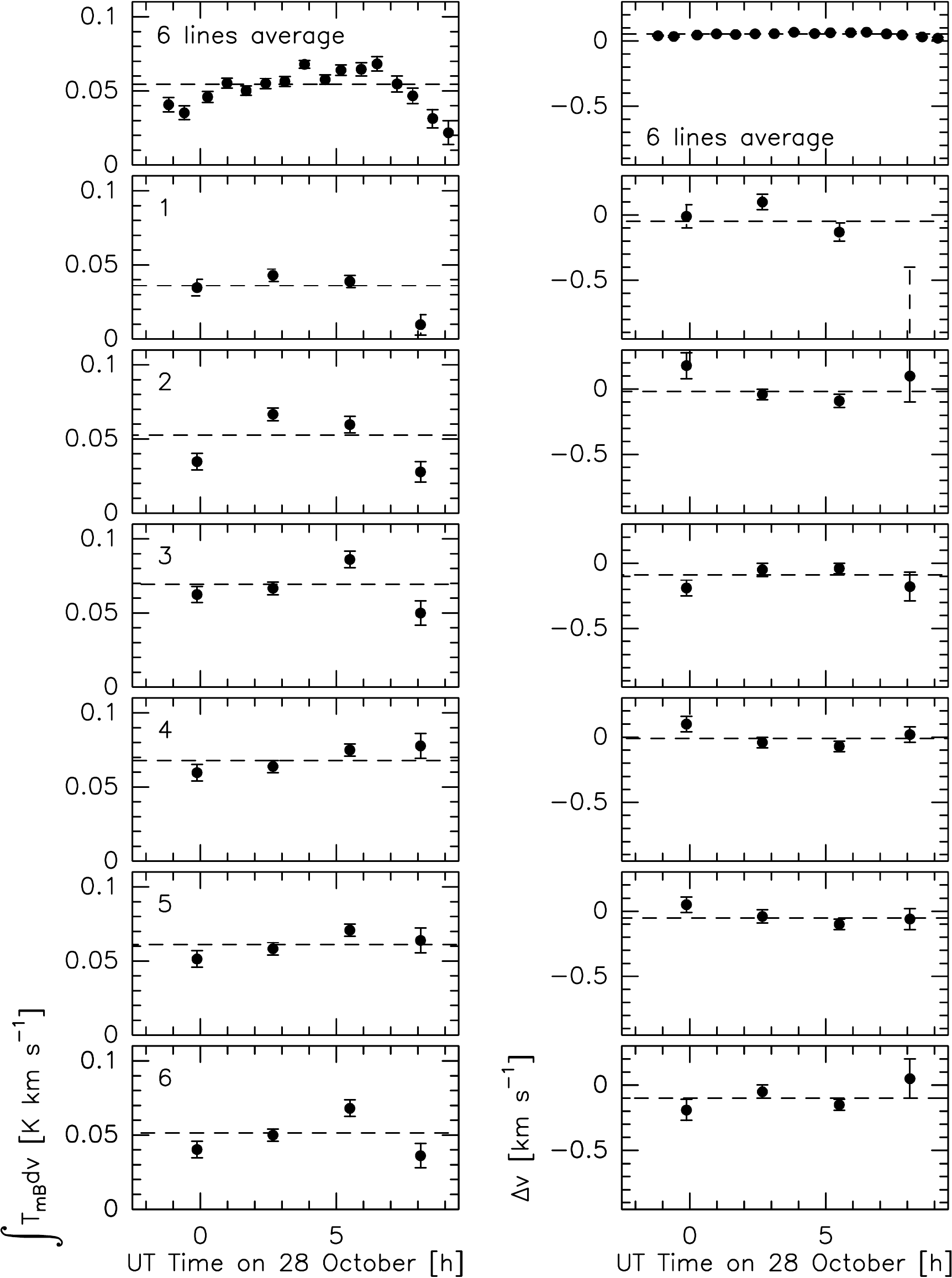}
\caption{Time evolution of the line area and Doppler shift of the six CH$_3$OH 157.2 GHz lines (labels 1 to 6) observed in On-Off mode. 
Each data point corresponds to 8~min on source spread over about 2~h. 
We present in the first row the evolution measured on the profile obtained by averaging the 6 lines. 
In that case, each data point corresponds to an individual On-Off observation with an integration time of 2~min on source.
In each frame, the horizontal dashed line represents the time-averaged value, measured on the  spectrum obtained when summing all the spectra acquired during the observations.
 {\bf Left pannels:} integrated area, in the main beam temperature scale; the label of the lines is indicated in the top left corner (assignation given in Table~\ref{tab-fluxes-line}).
{\bf Right pannels:} Doppler shift in the nucleus rest frame. }
 \label{fig-evolspec-meth}
 \end{center}
 \end{figure}

The On-Off spectra of the six methanol lines, averaging all data acquired on 28 October, are shown in Fig.~\ref{fig-oo}.
Time variations of the line intensities,  as well as those of the  mean line  velocities (Doppler velocity shift $\Delta v = \frac{\sum T_i v_i}{\sum T_i}$, where $i$ is the channel number which brigthness temperature and velocity are $T_i$ and $v_i$, respectively), are presented in Fig.~\ref{fig-evolspec-meth}.
We present also the variations of the average of the six lines, which has a better signal-to-noise ratio and can be studied with a better time resolution. We include in our analysis  single-dish spectra of the CH$_3$OH 157.2 GHz lines observed simultaneously in dual polarization  with the IRAM 30-m telescope in Spain (Fig.~\ref{fig-oo}). 
The observations were carried out between 0.63 and 1.78~h~UT on 28 October 2010.

\begin{figure}
 \begin{center}
\includegraphics[angle=0,width=\columnwidth]{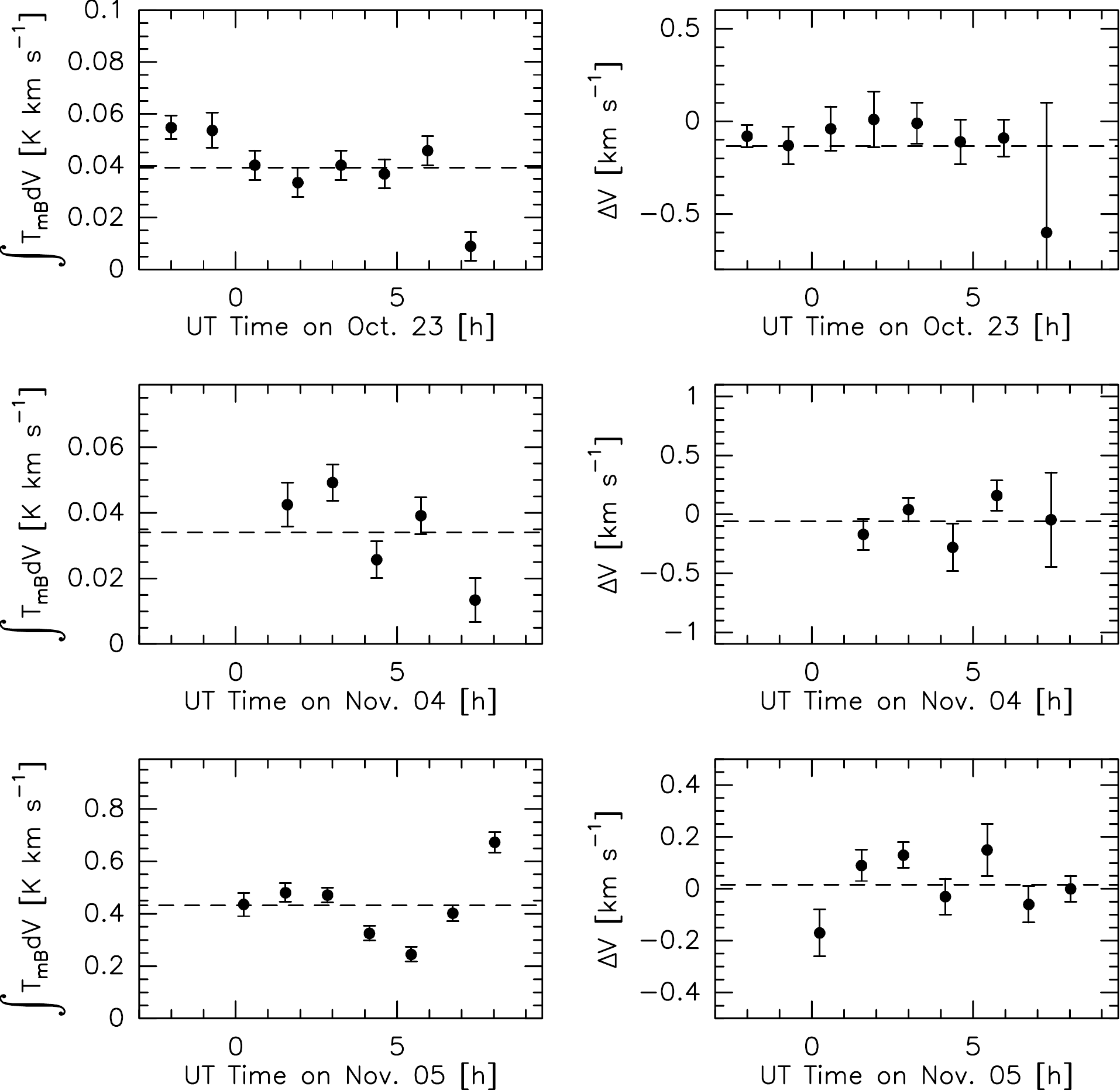}
\caption{Time evolution of the HCN lines observed in On-Off mode on 23 October (88.6 GHz, main hyperfine component, top frames), 4 November (88.6 GHz, main hyperfine component, middle frames), and 5 November (265.9 GHz, bottom frames).
 In each frame, the horizontal dashed line represents the time-averaged value, measured on the  spectrum obtained when summing all the spectra acquired during the observations.
 {\bf Left pannels:} integrated area (in main beam temperature scale); 
{\bf Right pannels:} time evolution of the Doppler shift of the lines.}
 \label{fig-evolspec-hcn}
 \end{center}
 \end{figure}

The time-averaged On-Off spectra of the HCN $J$(1--0) line obtained on 22.92--23.33 October and 4.04--4.33 November UT are presented in Figs~\ref{fig-oo}a and~\ref{fig-oo}b, respectively, whereas the time-averaged HCN $J$(3--2) spectrum obtained on 5 November is presented in Fig.~\ref{fig-oo}c. The time variations of the line integrated intensity and mean velocity are shown in Fig.~\ref{fig-evolspec-hcn}.

Table~\ref{tab-fluxes-line} lists the line-integrated intensities measured on the time-averaged HCN and CH$_3$OH spectra.

\begin{center}
\begin{table*}
\caption{Line intensities} 
\label{tab-fluxes-line}
\begin{tabular}{l  c c c c c c c}
\hline
\noalign{\smallskip}
\multicolumn{1}{l}{Date}  & \multicolumn{1}{c}{Molecule} & \multicolumn{1}{c}{Transition}  & \multicolumn{1}{c}{Freq.} & \multicolumn{1}{c}{RA$^a$} & \multicolumn{1}{c}{Dec$^a$}     & \multicolumn{1}{c}{Flux$^b$}       & \multicolumn{1}{c}{Beam}  \\
\multicolumn{1}{l}{UT}      & \multicolumn{1}{c}{}    &     \multicolumn{1}{c}{}        &\multicolumn{1}{c}{MHz}    & \multicolumn{1}{c}{$''$}     & \multicolumn{1}{c}{$''$}          & \multicolumn{1}{c}{Jy km s$^{-1}$} &\multicolumn{1}{c}{$''$}\\
\hline
\noalign{\smallskip}
\hline
\noalign{\smallskip}
\multicolumn{8}{l}{Plateau de Bure On-Off spectra}\\
\hline
\noalign{\smallskip}
27.956--28.379 Oct. & CH$_3$OH  & 1$_{0}$--1$_{-1}$ (1)$^c$   &157270.85 & -  & -&  0.68~$\pm$~0.05&  30.5    \\
27.956--28.379 Oct. & CH$_3$OH  & 2$_{0}$--2$_{-1}$ (2)$^c$   &157276.06 & -  & -&  0.99~$\pm$~0.05&  30.5    \\
27.956--28.379 Oct. & CH$_3$OH  & 3$_{0}$--3$_{-1}$ (3)$^c$   &157272.37 & -  & -&  1.30~$\pm$~0.05&  30.5    \\
27.956--28.379 Oct. & CH$_3$OH  & 4$_{0}$--4$_{-1}$ (4)$^c$   &157246.06 & -  & -&  1.27~$\pm$~0.05&  30.5    \\
27.956--28.379 Oct. & CH$_3$OH  & 5$_{0}$--5$_{-1}$ (5)$^c$   &157179.02 & -  & -&  1.14~$\pm$~0.05&  30.5    \\
27.956--28.379 Oct. & CH$_3$OH  & 6$_{0}$--6$_{-1}$ (6)$^c$   &157048.62 & -  & -&  0.96~$\pm$~0.05&  30.5    \\
22.904--23.321 Oct.  & HCN & (1--0) &\phantom{0}88631.85 & -  & -&   0.777~$\pm$~0.042&  54.0    \\
\phantom{0}4.042--4.342  \phantom{.}Nov.  & HCN & (1--0) & \phantom{0}88631.85 & -  & -&  0.666~$\pm$~0.049&  54.0    \\
\phantom{0}4.992--5.333  \phantom{.}Nov.  & HCN & (3--2) & 265886.43 & - & - & \phantom{0}7.49~$\pm$~0.22\phantom{0}&  18.0    \\
\hline
\noalign{\smallskip}
\multicolumn{8}{l}{IRAM 30-m spectra}\\
\hline
\noalign{\smallskip}
28.025--28.075 Oct. & CH$_3$OH  & 1$_{0}$--1$_{-1}$    &157270.85 &- &-&   0.382~$\pm$~0.066&  15.3         \\
28.025--28.075 Oct. & CH$_3$OH  & 2$_{0}$--2$_{-1}$    &157276.06 & - &- &  0.391~$\pm$~0.066&  15.3    \\
28.025--28.075 Oct. & CH$_3$OH  & 3$_{0}$--3$_{-1}$    &157272.37 & - &- &  0.566~$\pm$~0.066&  15.3   \\
28.025--28.075 Oct. & CH$_3$OH  & 4$_{0}$--4$_{-1}$    &157246.06 & - &- &  0.485~$\pm$~0.075&  15.3   \\
28.025--28.075 Oct. & CH$_3$OH  & 5$_{0}$--5$_{-1}$    &157179.02 & - &- &  0.471~$\pm$~0.080&  15.3   \\
28.025--28.075 Oct. & CH$_3$OH  & 6$_{0}$--6$_{-1}$    &157048.62 & - &- &  0.377~$\pm$~0.071&  15.3   \\
\hline
\noalign{\smallskip}
\multicolumn{8}{l}{Point source fitted to the interferometric data}\\
\hline
\noalign{\smallskip}
28.104--28.396 Oct. & CH$_3$OH  & 1$_{0}$--1$_{-1}$    &157270.85 & \phantom{--}0.72~$\pm$~0.39    & \phantom{--}0.06~$\pm$~0.31 & 0.065~$\pm$~0.016  &    3.12~$\times$~2.59 \\
28.104--28.396 Oct.  & CH$_3$OH  & 2$_{0}$--2$_{-1}$    &157276.06& \phantom{--}0.11~$\pm$~0.24     & --1.67~$\pm$~0.19         &  0.107~$\pm$~0.016 &     3.12~$\times$~2.59   \\
28.104--28.396 Oct.  & CH$_3$OH  & 3$_{0}$--3$_{-1}$    &157272.37& --1.33~$\pm$~0.33                 & --0.88~$\pm$~0.27         &  0.077~$\pm$~0.016&     3.12~$\times$~2.59  \\
28.104--28.396 Oct.  & CH$_3$OH  & 4$_{0}$--4$_{-1}$    &157246.06& \phantom{--}1.26~$\pm$~0.26       & --1.80~$\pm$~0.21        &  0.097~$\pm$~0.016&     3.12~$\times$~2.59  \\
28.104--28.396 Oct.  & CH$_3$OH  & 5$_{0}$--5$_{-1}$    &157179.02& \phantom{--}0.22~$\pm$~0.27       & --1.29~$\pm$~0.21         &  0.096~$\pm$~0.018&     3.12~$\times$~2.58  \\
28.104--28.396 Oct.  & CH$_3$OH  & 6$_{0}$--6$_{-1}$    &157048.62& \phantom{--}0.20 ~$\pm$~0.39        & --1.86~$\pm$~0.31       &  0.065~$\pm$~0.018&     3.12~$\times$~2.58  \\
22.917--23.333 Oct.  & HCN & (1--0)       & \phantom{0}88631.85 & \phantom{--}0.40~$\pm$~0.54 &--3.98~$\pm$~0.46 &  0.0258~$\pm$~0.005 &     6.21~$\times$~4.89  \\
\phantom{0}4.042--4.354 \phantom{.}Nov.  & HCN & (1--0)        & \phantom{0}88631.85& \phantom{--}1.42~$\pm$~0.34  &--3.14~$\pm$~0.41 &  0.0256~$\pm$~0.006&   3.73~$\times$~2.81   \\
\phantom{0}5.021--5.354 \phantom{.}Nov.  & HCN & (3--2)         & 265886.43         &  \phantom{--}0.14~$\pm$~0.08  &--1.97~$\pm$~0.12 &\phantom{00}0.24~$\pm$~0.04\phantom{0}& 1.32~$\times$~0.93 \\ 
\hline
\noalign{\smallskip}
\end{tabular}

$^a$ Position of the point sources fitted to the visibilities with respect to the reference positions of the comet given by the ephemeris from Giorgini et al. (\citeyear{gior+12}) (see Table~\ref{tab-obs} for the coordinates).\\
$^b$ Flux integrated over the whole line. The error does not include uncertainties in the absolute flux calibration:
10\% in On-Off, and 5, 10 and 15\% in interferometric mode at 88.6, 157.2, and 265.9 GHz, respectively.
For HCN $J$(1--0), the intensity is that of the main hyperfine component $F$2--1.
For HCN $J$(3--2), the intensity does not include the contribution from the faint (3.7\%) hyperfine components at --2.35 and 1.75 km s$^{-1}$.\\
$^c$ Label of the line used throughout the paper.
\end{table*}
\end{center}

\subsection{Interferometric maps}

\subsubsection{Spectral lines}
To produce interferometric maps of the methanol lines,
we averaged the channels in the spectral region  where emission is present in the On-Off spectra, i.e., over a width of 2.7~km~s$^{-1}$ for lines 1 to 4, and of 2.6 km s$^{-1}$ for lines 5 and 6 (the lines are labeled by their $J$ quantum numbers as indicated in Table~\ref{tab-fluxes-line}).
As cometary gas emission is extended, it is expected to be strongest on the shortest baselines \citep{boi+07}. Using coma parameters  deduced from On-Off spectra (CH$_3$OH production rate, gas temperature and velocity, see Sect.~\ref{sec-resu-line}), we computed the visibilities expected for the methanol lines using the model described in Sect.~\ref{sec-temp}. 
A comparison of the results with the noise level in the data showed that detection with baselines longer than 100~m was unlikely. 
Therefore, in order to improve the signal-to-noise ratio, we only considered baselines shorter than 100~m.
 As a result, the angular resolution of the presented maps is degraded by about 50\% with respect to what it would be using all baselines.
To measure the interferometric fluxes (refered as to $F_{\rm int}$ in Sect.~\ref{sec-radext}), we fitted  a point source model to the visibilities in the Fourier plane, prior to the imaging process (which introduces biases due to partial coverage of the $uv$ plane). 
The output of the fit is the position of the maximum line brightness (in the interferometric image) and the interferometric flux at this position. 
The methanol lines are detected with signal-to-noise ratios ranging from 3.6 to 6.8 (Table~\ref{tab-fluxes-line}).

Figure~\ref{fig-maps-hcn} presents  maps of the HCN $J$(1--0) and $J$(3--2) lines  integrated over a velocity range of 2.3 and 2.2 km s$^{-1}$, respectively. 
  Unlike methanol lines, all baselines were considered in the reduction process.
As for the methanol lines, we performed a fit of a point source to the visibilities to measure the line integrated intensities that are reported in Table~\ref{tab-fluxes-line}.

Spectra of the CH$_3$OH and HCN lines obtained in interferometric mode are shown in Fig.~\ref{fig-int}.

 \begin{figure}[h]
 \begin{center}
\includegraphics[width=\columnwidth]{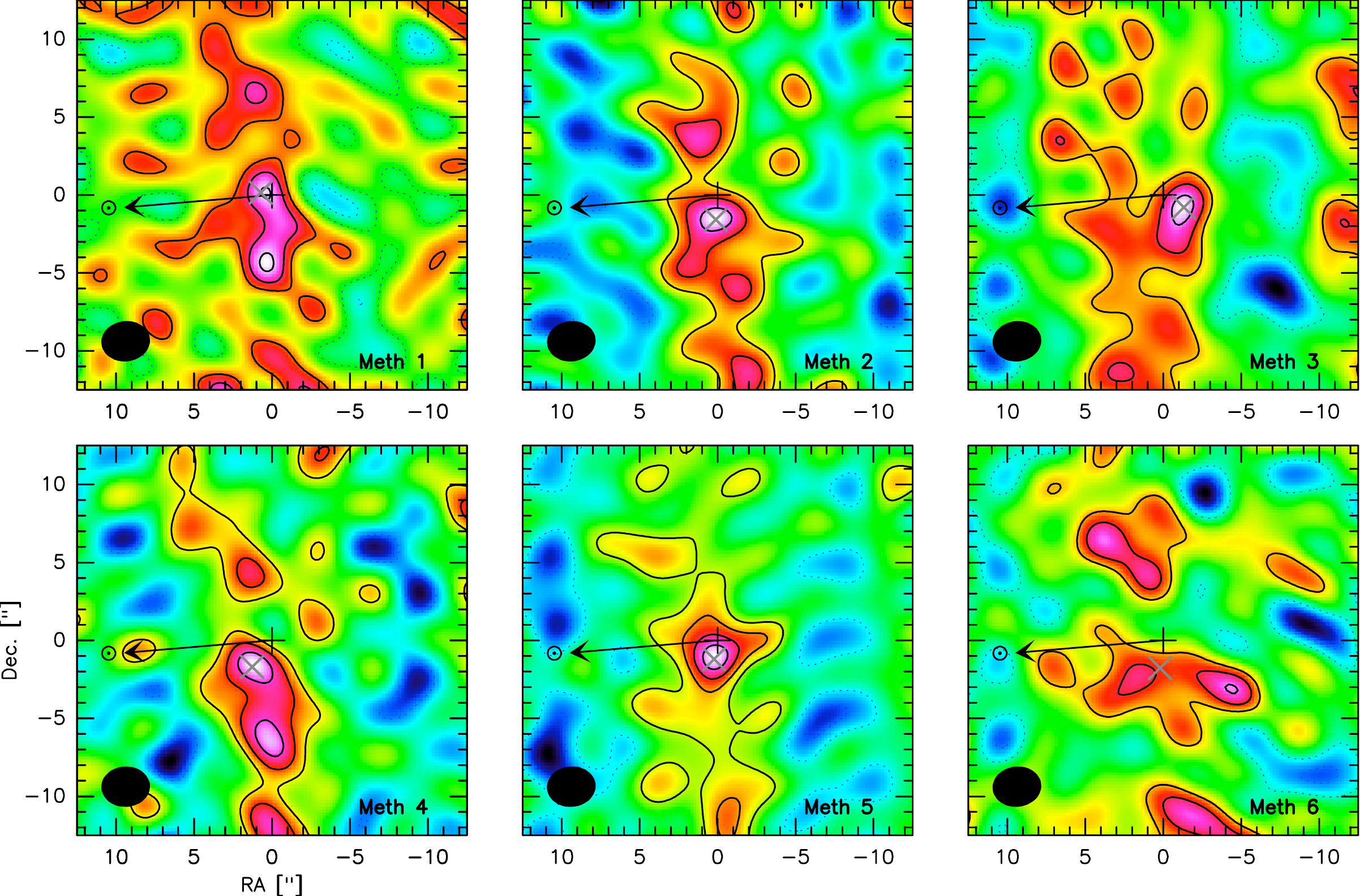}
 \caption{Interferometric maps of the six methanol lines observed on 28 October 2010 with the Plateau de Bure interferometer. Upper panels: lines 1 to 3; lower panels: lines 4 to 6. 
The contour spacing is 1-$\sigma$, i.e., 0.019, 0.024, 0.024,  0.022, 0.021 and 0.018 mJy~km~s$^{-1}$ for  lines 1 to 6, respectively.
Solid (respectively dotted) line contours stand for positive (respectively negative) signal. In each frame, the arrow indicates the direction of the Sun. The center of the maps is the position provided by the most recent ephemeris solution (reconstructed from EPOXI spacecraft navigation data, Giorgini et al. \citeyear{gior+12}). The ellipse in the lower left corner of the frames shows the synthesized beam. The position of peak brightness determined by fitting a point source to the visibilities (results in Table~\ref{tab-fluxes-line}) is indicated by the grey cross.}
 \label{fig-maps-meth}
 \end{center}
 \end{figure}

\label{sec-int-hcn}
\begin{figure}
 \begin{center}
\includegraphics[width=\columnwidth]{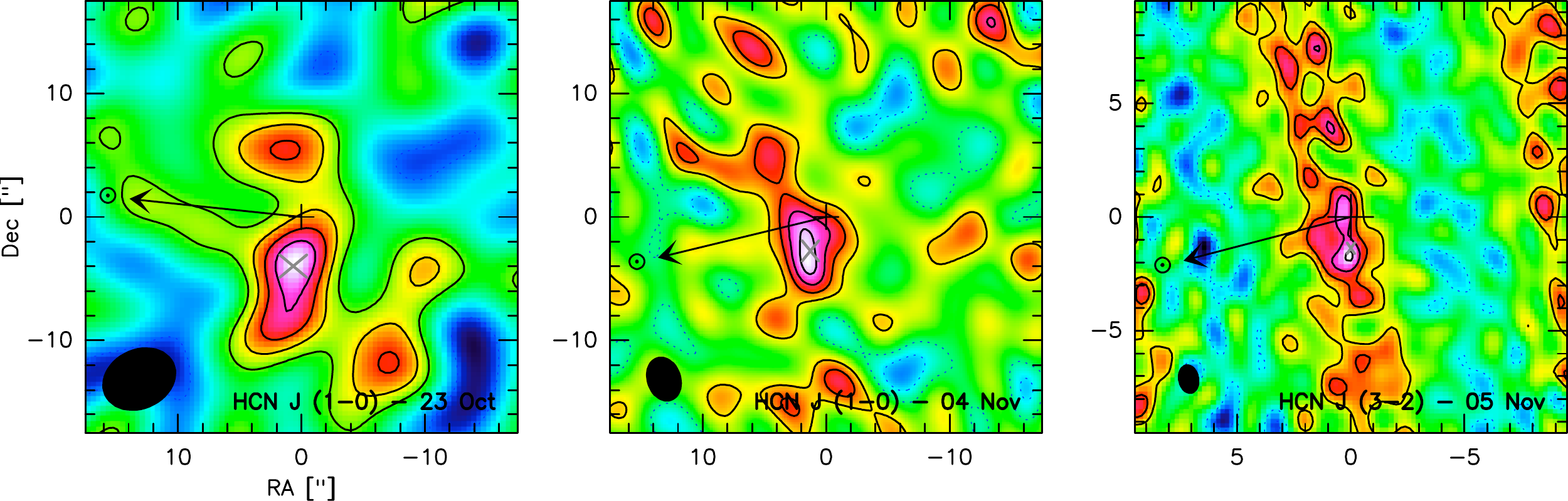}
 \caption{Interferometric maps of the $J$(1--0) and $J$(3--2) HCN lines observed with the Plateau de Bure interferometer. 
The contour spacing is 1-$\sigma$ (0.005, 0.006, and 0.064 Jy~km~s$^{-1}$ on 23 October 2010, 4 November 2010, and 5 November 2010, respectively). 
Solid (respectively dotted) line contours stand for positive (respectively negative) signal.
In each map, the arrow indicates the direction of the Sun. The cross at the center is the position provided by the most recentephemeris solution from Giorgini et al. (\citeyear{gior+12}) and  the ellipse in the lower left corner is the synthesized beam. 
The position of the point source fitted to the visibilities to measure the flux (Table~\ref{tab-fluxes-line}) is indicated by the superimposed grey cross.}
 \label{fig-maps-hcn}
 \end{center}
 \end{figure}

\begin{figure}
 \begin{center}
\includegraphics[angle=0,width=\columnwidth]{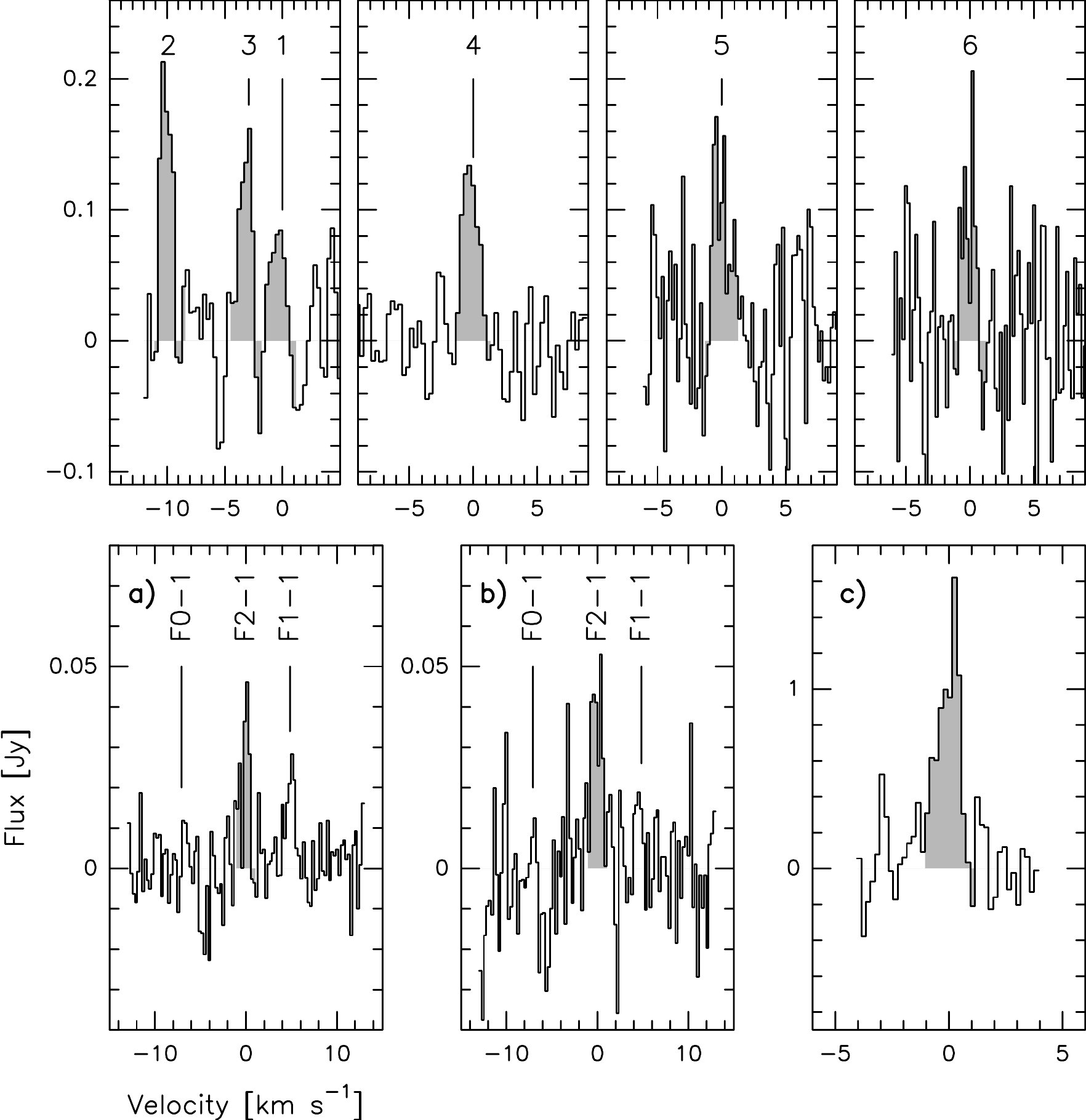} 
\caption{Interferometric spectra of comet 103P/Hartley 2. They were obtained by integrating the
signal in an ellipse of equivalent radius $r_e$ centred on ($RA_e$, $Dec_e$) with respect to the nucleus position given by the ephemeris.
{\bf Upper row:} Methanol 157.2 GHz lines on 28 October 2010, $r_e$ = 5$''$, ($RA_e$, $Dec_e$) = (--1.5$''$,--3$''$). The correspondence between labels and transitions is given in Table~\ref{tab-fluxes-line}. {\bf Lower row:} {\bf a, b)} HCN $J$(1--0) line on 23 October and 4 November 2010, respectively, with $r_e$ = 6.5$''$ and 4.5 $''$, and 
($RA_e$, $Dec_e$) = (1$''$,--5.5$''$) and (1.5$''$,--2.5$''$), respectively 
; {\bf c)} HCN $J$(3--2) line on 5 November 2010 with $r_e$ = 2.4$''$, ($RA_e$, $Dec_e$) = (0.5$''$,--1$''$).
The position of the hyperfine components of the HCN $J$(1--0) line is indicated in
frames {\bf a)} and {\bf b)}.}

\label{fig-int}
 \end{center}
 \end{figure}

\label{sec-int-cont}
\begin{figure}
 \begin{center}
\includegraphics[width=\columnwidth]{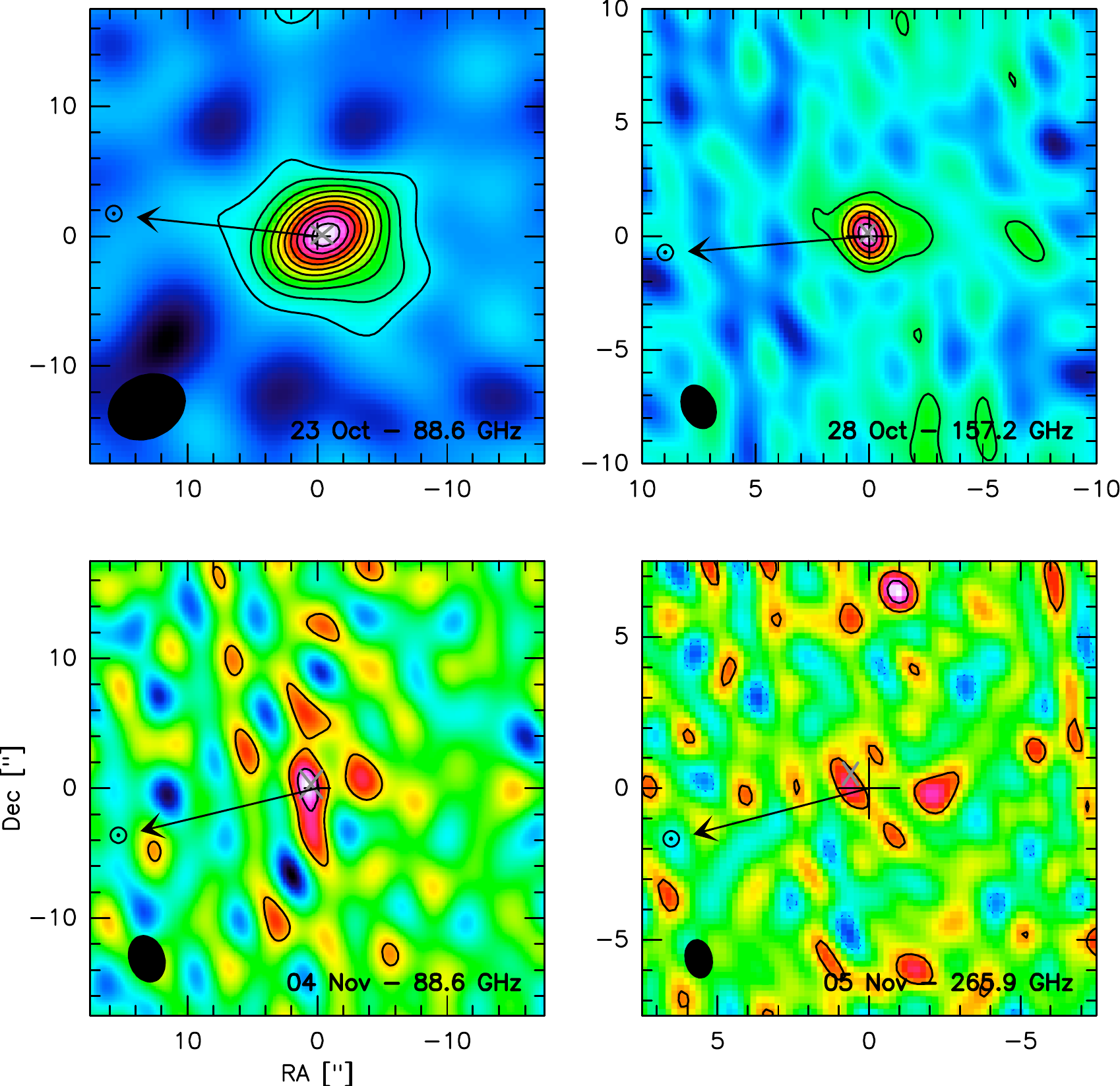}
 \caption{Interferometric maps of the continuum emission  from comet 103P/Hartley 2 observed with the Plateau de Bure interferometer. The dates and observing frequencies are indicated in the lower right corners of the frames. The contour spacing is 1--$\sigma$ (0.5 mJy) for the 265.9~GHz map (5 November) and 2--$\sigma$ for the other maps (0.06, 0.2, and 0.1~mJy for 23 October (88.6 GHz), 28 October  (157.2 GHz) and 4 November (88.6 GHz), respectively). In each map, the position of the point source fitted to visibilities is indicated by a  grey cross. The arrow indicates the direction of the Sun, the black cross at the center is the comet position provided by the ephemeris from Giorgini et al. (\citeyear{gior+12}), and the ellipse in the lower left corner is the synthesized beam. 
}
 \label{fig-cont}
 \end{center}
\end{figure}

\subsubsection{Continuum}

The continuum emission of the comet was estimated by averaging the signal in the line-free channels.
The corresponding maps are presented in Fig.~\ref{fig-cont}.
 The signal recorded for all baselines was used.
The results of the fits to a point-source are summarized in Table~\ref{tab-fluxes-cont}.

\begin{figure}
 \begin{center}
\includegraphics[width=\columnwidth]{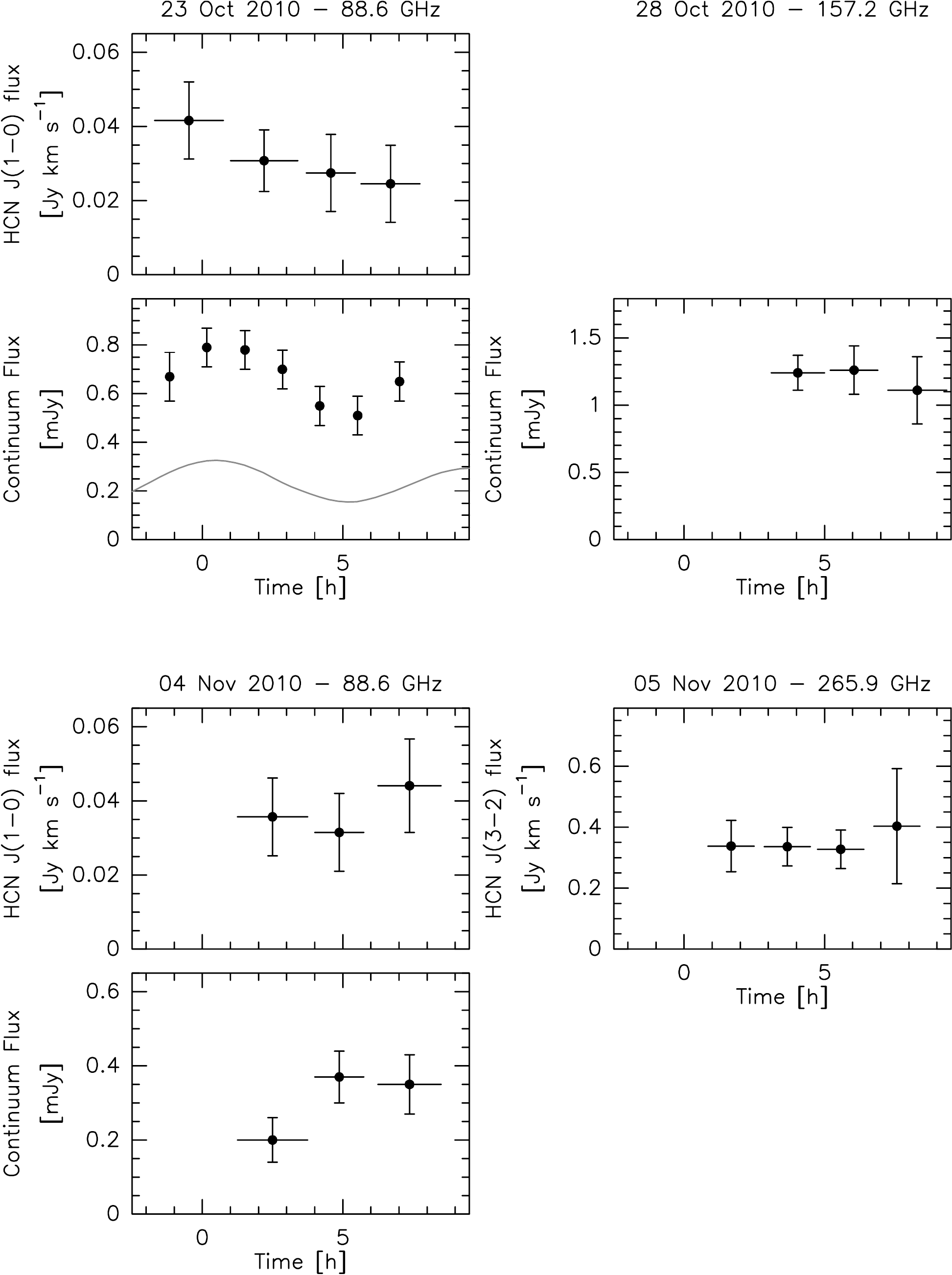}
 \caption{Time variations of the intensity of HCN lines and of continuum emission measured in 103P/Hartley~2 in interferometric mode.
On 23 October, the  sine-wave curve represents the synthetic  thermal emission curve of the elongated comet nucleus (see Sect.~\ref{sec-res-cont}). 
}
 \label{fig-evol-int}
 \end{center}
 \end{figure}

For HCN and the continuum, when possible, we divided the data into a number of subsets to study the time variations of the flux 
along the observing sequence, as shown in Fig.~\ref{fig-evol-int}. 

\section{Analysis of continuum data}
\label{sec-res-cont}
\subsection{Nucleus thermal emission}

When a comparison is possible (23 October 2010
and 4 November 2010) there is no obvious correlation between 
continuum and gas time variations (interferometric or On-Off data, see Fig.~\ref{fig-evol-int}). The continuum light curve observed on 23 October follows a sine-wave like variation, with the maximum and minimum separated by
$\sim$4.5 h, i.e., a quarter of the $\sim$18.3 h rotation period of the nucleus
\citep{har+11rad,mee+11}. This is unlike the light curves observed
for dust and gas phase species, which show rotation-induced variations with minima
and maxima separated by about half the rotation period 
\citep[][Sect.~\ref{sec-resu-line}]{ahe+11,dra+11,biv+11dps,besse+12}. 
Therefore, the variability of the continuum flux observed at
Plateau de Bure on 23 October suggests that the nucleus is the main source of the signal. The expected
thermal flux from the nucleus, assuming a spherical nucleus with a mean diameter
of 1.16 km \citep{ahe+11} and a simple Standard Thermal Model with $\eta$=1
\citep{lebspe89}, is given in Table~\ref{tab-fluxes-cont} for the different dates and
frequencies. The nucleus contributes from 30\% (23 October) to 55\% (5 November) of the observed continuum
fluxes. However, this approach does not take into account the fact that
the nucleus is elongated. 
 We thus calculated the nucleus synthetic thermal light curve for 23 October 2010, taking into account its shape (Thomas et al., 2013) and again a Standard Thermal Model with $\eta$=1. In the calculation we assumed a simple rotation at a constant rate, adopting the pole orientation RA=17\tdeg, Dec=+47\tdeg{} \citep{ahe+11,har+11rad} and the rotation period P=18.3~h \citep{thomas+12}, both measured at the epoch of the EPOXI flyby.
The results are given in Fig.~\ref{fig-evol-int}. 
 Since the rotation was excited and rapidly decelerating \citep[][and references therein]{belton+12},
the phasing of the thermal curve cannot be 
extrapolated over 12 days from 4 November 2010 to 23 October 2010, and 
we had to shift it by --6.5 h (or +2.5 h) to phase it with the 
observations of 23 October 2010. The synthetic
thermal light curve demonstrates that the temporal variations of the
continuum are indeed driven by the nucleus, and consistent with a rotation period of
about 18 h.

\subsection{Dust production rates}
\label{sec:dust}

The dust continuum flux was obtained by subtracting the nucleus thermal fluxes given 
in Table~\ref{tab-fluxes-cont}. To interpret the measurements, we used the dust model described by \citet{boc+10hst}.  
Dust thermal emission is modelled using absorption coefficients computed with the Mie theory  \citep[see, e.g.,][]{jewluu92,boi+12}. Optical constants pertaining to porous grains made of amorphous and crystalline silicates are calculated using the Maxwell Garnett mixing rule, adopting the refractive indices of astronomical silicates and forsterite published by \citet{dra85} and \citet{fab+01}, respectively. We assumed a porosity of 0.5, a crystalline/amorphous silicate ratio of 1 and a grain density $\rho_d$ = 500 kg m$^{-3}$. The equilibrium temperature of the grains contributing to the observed continuum emission was estimated as 275 K. We considered a differential dust production rate as a function of size $Q_d(a) \propto a^{-\alpha}$, with sizes between 0.1 $\mu$m and $a_{\rm max}$. Both radar images and images from the EPOXI mission showed that the grain size distribution in 103P/Hartley 2 coma included a significant population at decimeter 
sizes or larger \citep{har+11rad,ahe+11,kel+12}. Following \citet{crirod97}, we computed that the maximum liftable size through H$_2$O loading is $a_{\rm max}$ = 1 m, for a nucleus density $\rho_N$ = 500 kg m$^{-3}$ and isotropic outgassing with a water production rate $Q_{\rm H_2O}$ = 10$^{28}$ s$^{-1}$ (model 1). The EPOXI mission revealed strong enhancements of CO$_2$, dust and ice grain production from the smaller lobe of the nucleus \citep{ahe+11}. In addition there are several pieces of evidence that water was subliming from the icy grain halo \citep[e.g.,][see also Sect.~\ref{sec:5.3}]{mee+11,kni+12}. Therefore, we also investigated the case where the particles are ejected by CO$_2$ escaping gases. We assumed a surface gas mass flux corresponding to a total production rate $Q_{\rm CO_2}$ = 2 $\times$ 10$^{27}$ s$^{-1}$ in a cone of 90$^{\circ}$ opening angle (model 2).  This CO$_2$ production rate corresponds to the maximum value measured by EPOXI \citep{ahe+11,besse+12}. The opening angle of the cone is consistent with the spatial distribution of the large dust particles detected in radar echoes \citep{har+11rad}. We infer $a_{\rm max}$ = 2.4 m in this case. For comparison, size estimates of the large icy boulders detected by EPOXI imply radii $\leq$ 2.1 m \citep{kel+12}. The dust opacities link the observed continuum flux to the dust mass within the beam \citep{jewluu92}. For a size index $\alpha$ = --3.5 and $a_{\rm max}$ = 1 m (model 1), we get 7.8, 7.5 and 5.1 ($\times$ 10$^{-3}$) kg m$^{-2}$ at 1.2, 1.5 and 3.4~mm, respectively. For $a_{\rm max}$ = 2.4 m (model 2), the inferred dust opacities are 3.9, 3.7 and 2.7 ($\times$ 10$^{-3}$) kg m$^{-2}$, respectively. We assumed a nominal value $\alpha$ = --3.5 in the dust production rate calculations. Indeed, the size index in the coma $q$ is related to the size index at production $\alpha$ through $q$ = $\alpha$ + 0.5, for a velocity law in the form $a^{-0.5}$ \citep{fulle+04}. \citet{bau+11} inferred $q$ = --3.0 $\pm$ 0.3 (corresponding to $\alpha$ = --3.5) from $WISE$ observations of 103P/Hartley 2 undertaken in May 2010.

An important parameter for determining the dust production rate is the dust velocity which is a function of the grain size and surface gas production. 
For model 1, the size-dependent grain terminal velocities deduced from the formula given by \citet{crirod97} vary from 0.45 to 454 m s$^{-1}$ in the size range 0.1~$\mu$m--1m, and follow approximately a size dependence  $a^{-0.5}$. For model 2, they range from 0.74 to 560 m s$^{-1}$ for $a$ = 0.1~$\mu$m--2.4 m. We checked that the velocities of large-size grains ($>$ 1~mm) are in agreement with state-of-the-art dusty gas dynamic modelling of 103P/Hartley 2 under similar assumptions (V. Zakharov, private communication). We expect models 1 and 2 to bracket the actual kinematic properties of the dust grains at the time of the Plateau de Bure observations, 
as they correspond to two extreme cases of outgassing profiles. However, the velocities derived for 2-cm grains are 3.2 m s$^{-1}$ and 7.9 m s$^{-1}$, for models 1 and 2 respectively, and much lower than the value of 30 m s$^{-1}$ estimated for this size from the Doppler spectrum of the radar echo \citep{har+11rad}. Accelerating centimeter size grains to this velocity require very large surface gas fluxes ($\sim$ 4 $10^{-4}$ g cm$^{-2}$ s$^{-1}$), implying highly collimated gas and dust production. Dust production rates were estimated using the dust velocities given by our models. Higher production rates would be obtained using instead velocities measured by the radar echoes.

For model 1, averaging values obtained for the four observing dates, we infer a mean dust production rate of 830 kg s$^{-1}$ for a size index $\alpha$ = --3.5. The dust production rates obtained for 23, 28 October, 4 November and 5 November are 1000, 690, 740, and 900 kg s$^{-1}$, respectively (Table~\ref{tab-fluxes-cont}). These values are 2--3 times higher than the estimates obtained by  \citet{har+11rad} from coma radar echoes using the same size index. Dust production rates obtained using model 2 are larger than for model 1, as both the grain maximum size and velocities are larger: the mean value for the observing period is 2700 kg s$^{-1}$ , with a value of 2300 kg s$^{-1}$ for 4 November. We note that, for $\alpha$ = --4, the two data sets at 3.4~mm yield  higher production rates (by a factor of 2) than 1.2 and 1.5~mm data, suggesting that such a steep size distribution can be excluded. Tighter constraints on the size distribution cannot be obtained from the present data set since the data were acquired
  on different dates. The masses derived using $\alpha$ = --3.0 are 50--70\% higher than values derived with $\alpha$ = --3.5. The two data sets at 3.4~mm suggest that the dust production rate was on average 35\% higher on 23 October than on 4 November.            

In summary, uncertainties in the size distribution cut-off and kinematics make our estimate of the dust production rate quite uncertain. The large dust velocities measured by the radar measurements favor models leading to dust production rates higher than the range of values 830--2700 kg given by models 1 and 2. These values are larger than the CO$_2$ and H$_2$O production rates of 160 kg s$^{-1}$ and $\sim$ 300 kg s$^{-1}$, respectively, measured in the period of the Plateau de Bure observations \citep{ahe+11,del+11,cro+12}. They imply a dust-to-gas production rate ratio in the range 2--6, suggesting that 103P/Hartley 2 was highly dust productive. This large dust-to-gas ratio can be explained by the unusual  activity of the comet given its size, which allows decimeter size particles or larger boulders to be entrained by the gas due to the small nucleus gravity.

\begin{table*}
\begin{center}
\caption{Characteristics of continuum maps, nucleus contribution, and dust production rates.}
\label{tab-fluxes-cont}
\begin{tabular}{l  c c c c  c c}
\hline
\noalign{\smallskip}
Mean date & Frequency & RA$^a$ & Dec$^a$ & Observed Flux & Nucleus$^b$  & Dust production rates  \\
UT & MHz       & $''$     & $''$      & mJy           &  mJy                  & kg s$^{-1}$   \\
\hline
\noalign{\smallskip}
\hline
\noalign{\smallskip}
 23.125 Oct.         & \phantom{0}88631.847 &  --0.64~$\pm$~0.13            & \phantom{--}0.14~$\pm$~0.11 &  0.65~$\pm$~0.03 & 0.19 & (1000$^c$, 3100$^d$)\phantom{00}    \\ 
 28.250 Oct.          &  157172.000          &   \phantom{--}0.05~$\pm$~0.05 &\phantom{--}0.05~$\pm$~0.06  &  1.23~$\pm$~0.09  & 0.52 & (690$^c$, 2270$^d$)\phantom{0}    \\
 \phantom{0}4.196 Nov. & \phantom{0}88631.847 &    \phantom{--}0.62~$\pm$~0.22&--0.09~$\pm$~0.26  &  0.25~$\pm$~0.04 & 0.10 & (740$^c$, 2300$^d$)\phantom{0}    \\
 \phantom{0}5.168 Nov.  & 265886.432           &    \phantom{--}0.76~$\pm$~0.12&\phantom{--}0.08~$\pm$~0.18 &  1.69~$\pm$~0.48 & 0.95 & (900$^c$, 2970$^d$)\phantom{0}    \\
\hline
\noalign{\smallskip}
\end{tabular}
\end{center}
$^a$~Position of the fitted point source with respect to the comet ephemeris from Giorgini et al. (\citeyear{gior+12}) (see Table~\ref{tab-fluxes-line}).
$^b$~Nucleus contribution computed  for a mean radius of 1.16 km \citep{ahe+11}.
$^c$~Dust production rates computed using parameters of model 1 (see text).
$^d$~Dust production rates computed using parameters of model 2 (see text). 
\end{table*}

\section{Analysis of spectral data}
\label{sec-resu-line}

\subsection{Coma temperature}
\label{sec-temp}

The gas temperature in cometary atmospheres is an important parameter for computing 
reliably the rotational excitation of the  molecules.
Rotational temperatures can be measured from the intensities of several rotational lines of a given molecule which are, under optically thin conditions, proportional to
the column densities within the upper transition states. Assuming that the population distribution can be described by a single rotational temperature $T_{rot}$,
then the relation between line intensities and the energy of the upper transition states
is directly related to $T_{rot}$ 
(see \citet{gollan99} and \citet{boc+94aa}  for  detailed descriptions of the rotation-diagram technique).

\begin{figure}
 \begin{center}
\includegraphics[width=\columnwidth]{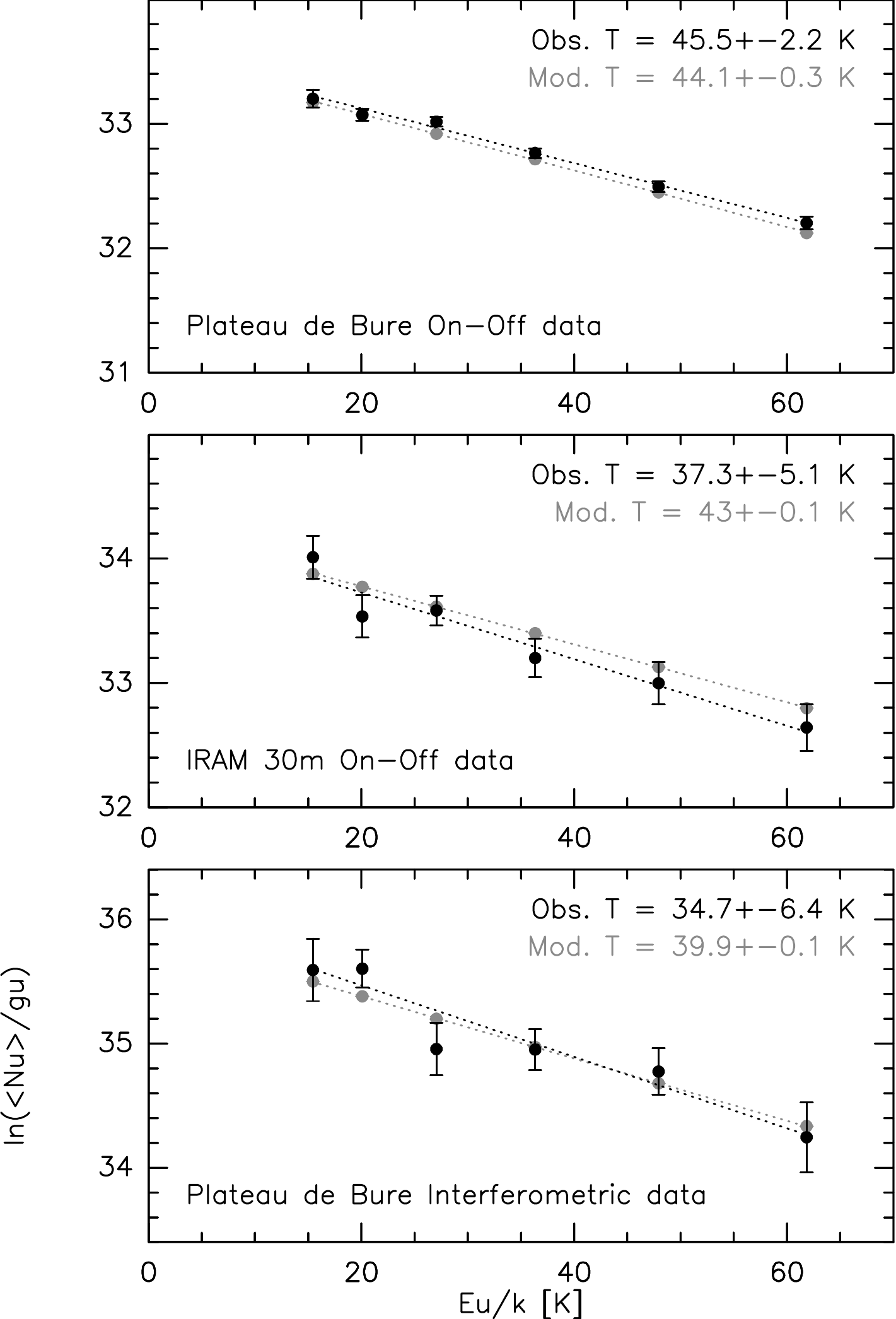}
 \caption{Rotation diagrams deduced from the methanol observations of 28 October 2010  (black symbol) and simulations  (grey symbols) for On-Off observations with the Plateau de Bure (top frame), IRAM 30-m observations (middle frame) and interferometric observations (lower frame). The corresponding beam sizes are 30.5$''$, 15.3$''$, and 2.8$''$, respectively, and the projected beam radii are $\sim$ 1500, 780, and 150 km, respectively. The model assumes a purely nuclear source for methanol and a constant gas temperature of 40~K. The rotational temperatures (with 1$\sigma$ uncertainty) deduced from the rotation diagrams are given. }
 \label{fig-rd}
 \end{center}
\end{figure}

Figure~\ref{fig-rd} shows rotation diagrams with the inferred $T_{rot}$ obtained from the analysis of the time-averaged CH$_3$OH data acquired on 27--28 October with the Plateau de Bure (interferometric and On-Off modes) and the IRAM 30-m telescopes.
Values range from 35 to 45 K. As observed methanol lines are all of E-type, the derived rotational temperatures are not affected by any potential E/A anomaly.

The signal-to-noise ratio of the Plateau de Bure On-Off data allows us to measure $T_{rot}$ for the first three out of the four data subsets shown in Fig.~\ref{fig-evolspec-meth}, to study temporal variations. 
We infer temperatures of $48.0 \pm 8.3$~K at 23.9~h UT on 27 October,  and $35.4 \pm 3.1$ K and $53.3 \pm 6.5$ K
at 2.8 and 5.5~h UT on 28 October. 
The signal-to-noise ratio in the last data subset is too low to infer a reliable rotational temperature from the rotation diagram.
Temporal variations of the temperature are observed, with a minimum at  $\sim$ 2.8~h~UT, about 15~K lower than values for 27 October at 23.9~h~UT and 28 October at 5.5~h~UT. 
\citet{dra+12} monitored 157.2 GHz methanol lines with the IRAM 30-m telescope and derived even stronger temperature variations. Higher temperatures are measured at times of maximum CH$_3$OH production \citep{dra+12}, 
suggesting that the coma was non-isothermal with a higher temperature in the gas jet (emanating from the small end of the nucleus, see Sect.~\ref{sec-jet}). 
This temperature enhancement might be related to gas release from icy grains, as modelled by  \citet{fougere+12}. \citet{bonev+12} have suggested that the pronounced asymmetries observed for the rotational temperature of H$_2$O reflect different mechanisms of water release.

\begin{figure}
 \begin{center}
\includegraphics[width=\columnwidth]{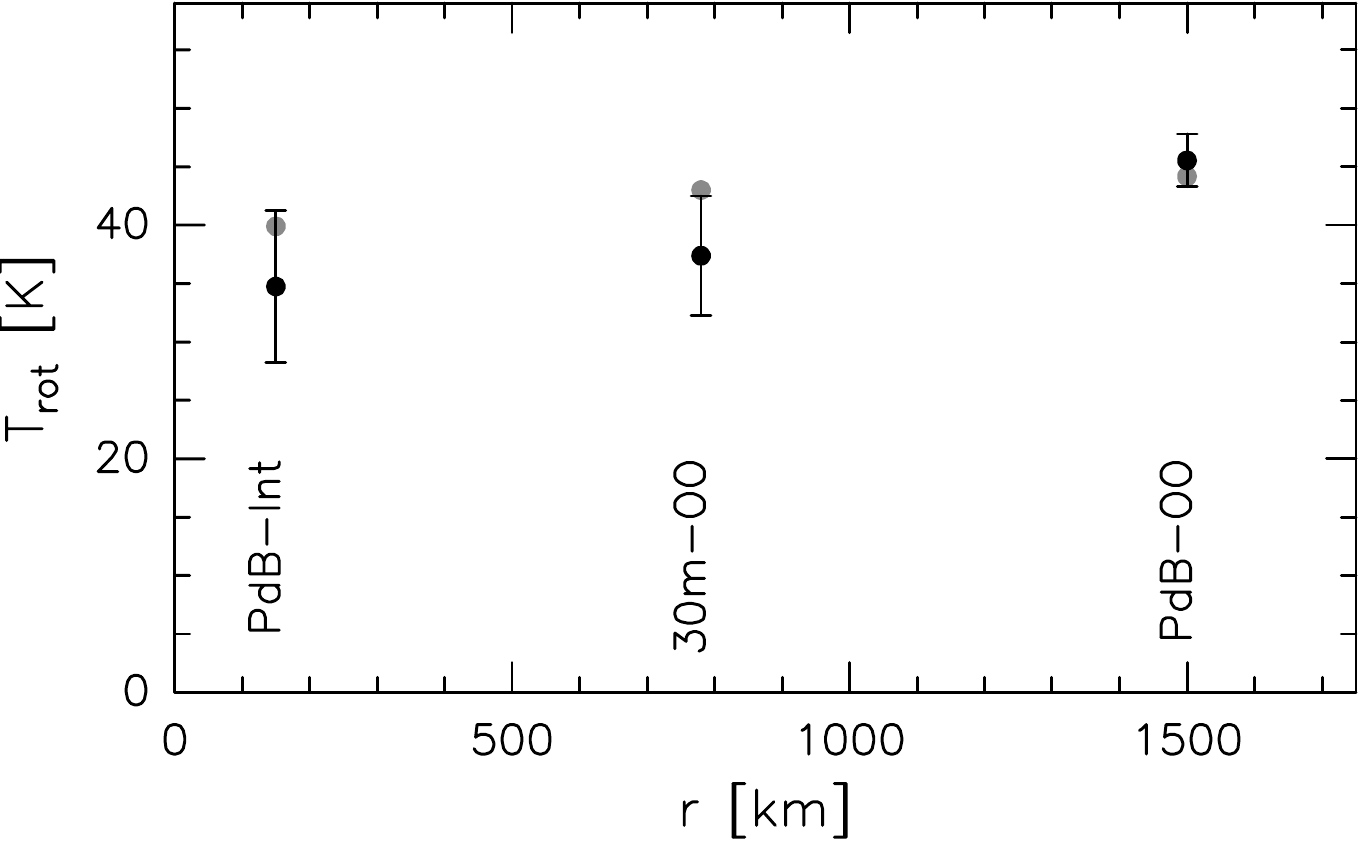}
 \caption{CH$_3$OH rotational temperature as a function of beam radius (black symbols). The labels PdB-Int, PdB-OO, and 30m-OO refer to temperature measurements from observations with the Plateau de Bure interferometer in interferometric and  On-Off modes, and with the IRAM 30-m telescope, respectively.
Results obtained from a model with a constant gas temperature of 40~K are overplotted in grey. }
 \label{fig-profT}
 \end{center}
\end{figure}

Figure~\ref{fig-profT} shows rotational temperatures as a function of projected beam radius ($\sim$150 and $\sim$1500~km projected on the comet). A slight increase with increasing beam radius is suggested, which could result from increased efficiency of photolytic heating, as expected from coma thermodynamic models  \citep{com+04}. Alternatively, the increase of the rotational temperature of the 157.2 GHz methanol lines could be related to the breakdown of thermal equilibrium in the region of the coma where radiative processes such as radiative decay and infrared excitation become significant with respect to collisional excitation. 

In order to check this later hypothesis, synthetic On-Off spectra and visibilities were computed
using an excitation model and a radiative transfer code, following
\cite{boi+07}. We summarize here the main aspects of the computations.
The calculation of the population of the rotational levels of the
 molecules considers collisions with H$_2$O and electrons, and IR
radiative excitation. 
The model provides populations as a function of radial distance $r$, given
an H$_2$O density law with $r$. For simplicity, we assumed an isotropic
H$_2$O coma. The total cross-section of methanol-water collisions was taken equal to
$\sigma_c = 5\times 10^{-14}$ cm$^2$   \citep{biv+99}. 
The H$_2$O production rate was assumed to be 1.0 $\times$ 10$^{28}$ s$^{-1}$, i.e., within the range $(0.5-1.5) \times 10^{28}$ s$^{-1}$ of values measured during the observing period \citep{ahe+11,com+11,del+11,mum+11,cro+12}.
We considered outflow at a constant velocity of 0.7~km~s$^{-1}$, which corresponds to the half width at half maximum of the HCN $J$(1--0) line observed in On-Off mode, and is consistent with other determinations \citep{cro+12,biv+11dps}.
The brightness distribution of the  lines was then computed
using the radiative transfer code of \cite{boi+07}, which includes
both self-absorption and stimulated emission (see \citealt{biv97} for detailed explanations). 
On–-Off synthetic spectra are 
obtained by convolving the brightness distributions with the primary
beam pattern of the  antennas, described by a 2D Gaussian.
Visibilities are calculated from the Fourier transform of the
brightness distribution \citep{boi+07}. 
The modeled and observed interferometric flux densities, both obtained by  fitting a point source to the visibilities, can then be directly compared.

In Fig.~\ref{fig-rd} we compare the observed rotation diagrams with the results 
of a simulation with a constant (40~K) gas temperature in the coma,  and isotropic outgassing. 
The rotation diagrams derived from the Plateau de Bure data (On-Off and interferometry) 
are well reproduced. The agreement is not as good for the 30-m data but this can be 
explained by the time variations of the temperature: according to the Plateau de Bure On-Off data, the temperature was minimum at the time the 30-m telescope data were acquired.
As shown in Fig.~\ref{fig-profT}, the model predicts an increase of the methanol rotational temperature
with increasing beam size which is not as strong as observed but reproduces the data within 1$\sigma$.

Rotational temperatures as high as 70--90 K were derived from infrared observations  
with field of view of 50--100 km \citep{mum+11,del+11,bonev+12,kawa+13}. Since these measurements were performed on 22 October and 4 November, i.e., one week before and  after our observations, the factor of two difference with our measurements might be related to time variations of the mean temperature of the coma. A more likely explanation is a decrease of the gas kinetic temperature with increasing nucleocentric distance, as expected for adiabatic expansion of the gas. Indeed, \citet{bonev+12} report measurements of the 
spatial distribution of the H$_2$O rotational temperature showing a rapid decrease with increasing nucleocentric distance. We did not consider radial variations of the temperature 
to interpret the rotational temperature of methanol lines, because of the complexity of 
the coma.

\subsection{Gas production rates from single-dish data}
\label{sec-qp}

The production rates derived from the single-dish data were computed using the model described in the previous section and assuming an isotropic outflow and a nuclear origin for the molecules. The gas temperature was taken equal to 40~K and assumed to be constant in the coma, and the expansion velocity was taken equal to 0.7~km s$^{-1}$ (Sect.~\ref{sec-temp}). 

The production rates deduced from daily-average spectra are   $Q^{\rm CH_3OH}$ = (3.15 $\pm$ 0.17) $\times~10^{26}$~s$^{-1}$ for 28 October, and  $Q^{\rm HCN}$ = (1.4 $\pm$ 0.2), (1.4 $\pm$ 0.2), and (0.60 $\pm$ 0.07)  $\times~10^{25}$~s$^{-1}$ for 23 October, 4 November, and 5 November, respectively (the error bar only includes uncertainties on the line intensity).  Our production rate determinations are consistent with those derived by \citet{dra+12}. Detailed modelling considering jet activity and production from a distributed source would lead to higher production rates. 

Using water production rates published in the literature \citep{mum+11,del+11,biv+11dps,com+11,cro+12}, 
we derive average abundances relative to water of  0.16\% for HCN and 2.7\% for CH$_3$OH which are similar to abundances commonly measured in cometary atmospheres.
The HCN abundance is $\sim$30\% lower  than values derived by \cite{mum+11} and \cite{del+11} from infrared observations. The methanol abundance is consistent with the determination made by \cite{mum+11}, but is about twice higher than the value measured by \cite{del+11}.  However, \citet{Vil+12} derived a methanol abundance of 1.8\% from improved modelling of CH$_3$OH infrared emission. The discrepancy between radio and IR results is not surprising given the strong temporal variability of the coma and differences in field of view.

\subsection{Time variations and spatial distribution}
The time evolution of the intensity and mean velocity of molecular lines can reveal important information about nucleus rotation and outgassing properties, as illustrated by radio studies of, e.g., C/2001 Q4 (NEAT) and C/1995 O1 (Hale-Bopp) \citep{biv+09,boc+09hb}. Rotation-modulated variations of these two quantities were observed 
in 103P/Hartley 2 by \citet{dra+11,dra+12}, from an extensive monitoring of HCN and CH$_3$OH millimeter lines undertaken from the end of September to early December 2010. 
 Our data set provides complementary time coverage, as well as additional information through the position of the peak line brightness in the interferometic maps.

\subsubsection{Production rate and velocity shift}
 \label{line_I_V}

\begin{figure}
%  \begin{center}
%\includegraphics[angle=0,width=7cm]{fig-fluxdvpa-mod.eps}
\hspace{-1cm}\includegraphics[angle=0,width=10cm]{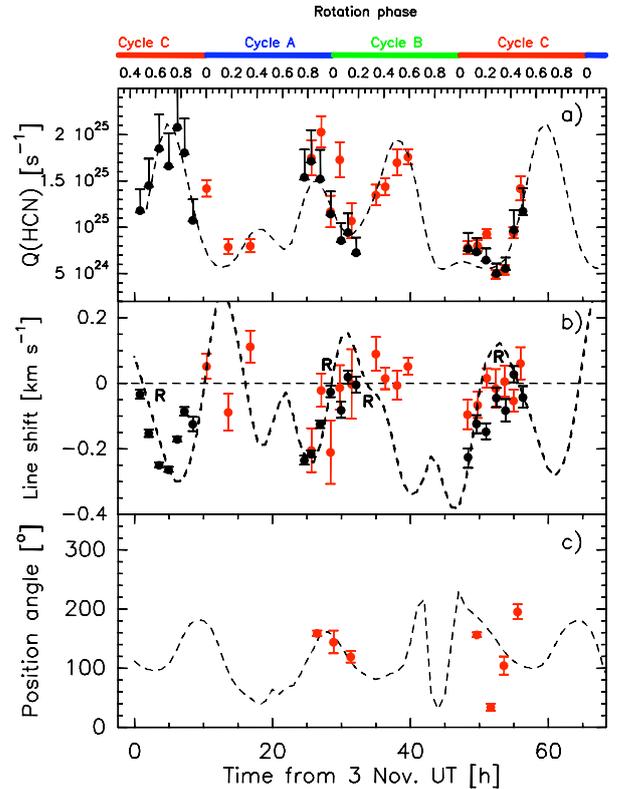}
\caption{Variation of the HCN production rate (a), velocity shift (b), and position angle of HCN peak brightness (c) with time. Data obtained with the Plateau de Bure interferometer and CSO are plotted with red symbols, whereas those from 
\citet{dra+12} are plotted with black symbols. For Plateau de Bure and CSO data, the line shift corresponds to the first moment of the line (Sect. 2.2.1), whereas the line shift published by \citet{dra+12} is the median velocity
of the spectrum. Simulations show that the two quantities are comparable. Spectra showing a single redshifted peak 
are indicated by the symbol "R". The dashed curves in the figures are respectively: a) the temporal variation of the insolation of the SL region covering $lat$ = 70--90$^{\circ}$, $long$ = 0--180$^{\circ}$ (Fig.~\ref{fig-section}), shifted by 1.71 h, to account for delayed HCN production (see text, Sect.~\ref{sec-jet}); b) for this region, 
the weighted mean of cos($\pi$--$Ap$) of insolated area; c) the weighted mean of the position angle of the normal to insolated area. The weight for surface elements are their area times the cosine of the solar zenith angle. Curves in (a) and (b) plots are in arbitrary units. The cycles indicated at the top are those defined by  \citet{dra+11}. The rotational phase of 0.5 corresponds to the time of EPOXI closest approach. The correspondence between phase and date was computed assuming a precession (also called rotation) period of 18.4 h.  }
 \label{fig-Qcurve1}
% \end{center}
 \end{figure}

\begin{figure}
  \begin{center}
\includegraphics[angle=-90,width=\columnwidth]{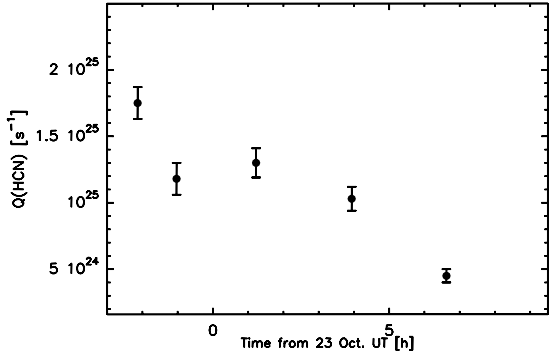}
\caption{HCN production rates on 22--23 October 2010 derived from the Plateau de
Bure On-off observations.
}
 \label{fig-Qcurve2}
 \end{center}
 \end{figure}

The evolution of the HCN production rate in the time range 3--6 November 2010 is shown in Fig.~\ref{fig-Qcurve1}a. 
The display combines values derived from the Plateau de Bure On-Off data, and from observations at the Caltech Submillimeter Observatory (CSO, Biver et al., in preparation), and IRAM 30-m telescope \citep{dra+12}. Production rates were computed as explained in Section~\ref{sec-qp}, except for the measurements from the IRAM 30-m telescope, where we used values published by \citet{dra+12}. On 4 November, the HCN production rate presented a minimum near 8 h UT and is near its maximum at the time of the EPOXI closest approach ($\sim$ 14 h UT). On 5 November, the intensity of the HCN line observed in On--Off mode decreased from 2 to 5 h UT and then displayed a sharp increase, a behavior consistent with measurements from \citet{dra+12}, reported in Fig.~\ref{fig-Qcurve1}. 
 HCN production rates deduced from the 22--23 October single-dish data are displayed in Fig.~\ref{fig-Qcurve2}.

We compared the time variations observed in interferometric (Fig.~\ref{fig-evol-int}) and single-dish On-Off mode (Figs.~\ref{fig-evolspec-meth} and \ref{fig-evolspec-hcn}), taking into account the different beam sizes. With respect to the interferometric mode, the response of the On-Off mode to an increase of gas production is delayed by $\Delta t = (r_{beam}-r_{int})/v_{exp}$, where $v_{exp}$ is the gas expansion velocity, and $r_{beam}$ and $r_{int}$ are the On-Off and interferometric beam radii, respectively.
Assuming $v_{exp}$ = 0.7~km s$^{-1}$ (Sect.~\ref{sec-temp}), we estimate that  $\Delta t$ is of the order of 20~min for the observations at 265.9 GHz on 5 November,  50~min for the observations at 88.6 GHz on 23 October, and 70~min for the observations at 88.6 GHz on 4 November. The same numbers are obtained using model simulations from \citet{biv+07}. Taking into account the beam delay, we found that time variations in the two modes are consistent. For exemple, on 4 November, the On-Off intensity of the HCN line shows a regular decrease from 1 to 8 h UT.  An increase of the interferometric flux from 5 to $\sim$ 7 h UT, not seen in the On-Off mode because of the time delay, is marginally observed (Fig.~\ref{fig-evol-int}). This trend is consistent with the increase of gas production rates observed between 10.8 and 15.9~h UT from infrared observations with the Keck Observatory \citep{del+11}.

The time variations of the velocity shift $\Delta v$ of HCN and CH$_3$OH lines observed in On-Off mode are presented in Figs~\ref{fig-evolspec-meth} and \ref{fig-evolspec-hcn}. Averaging daily data, the lines are slightly shifted towards negative velocities, consistent with excess gas production towards the Sun since the phase angle was $\phi$ $\sim$ 55$^{\circ}$. In Fig.~\ref{fig-Qcurve1}b, our HCN $\Delta v$ data for 4 and 5 November are combined with those of \citet{dra+12} measured with higher signal-to-noise ratios. Measurements at common times generally agree. Spectra with large blueshifts are characterized 
by the presence of a single blueshifted peak at about --0.6 km s$^{⁻1}$, whereas spectra with modest velocity shifts 
present blueshifted and redshifted peaks of similar intensities, or a redshifted peak at $\sim$ 0.3 km s$^{-1}$ \citep{dra+12}. For the purpose of discussion (Sect.~\ref{sec-jet}), data showing a well-defined redshifted peak are indicated in Fig.~\ref{fig-Qcurve1}b.

 \begin{figure} 
  \begin{center}
\includegraphics[angle=0,width=\columnwidth]{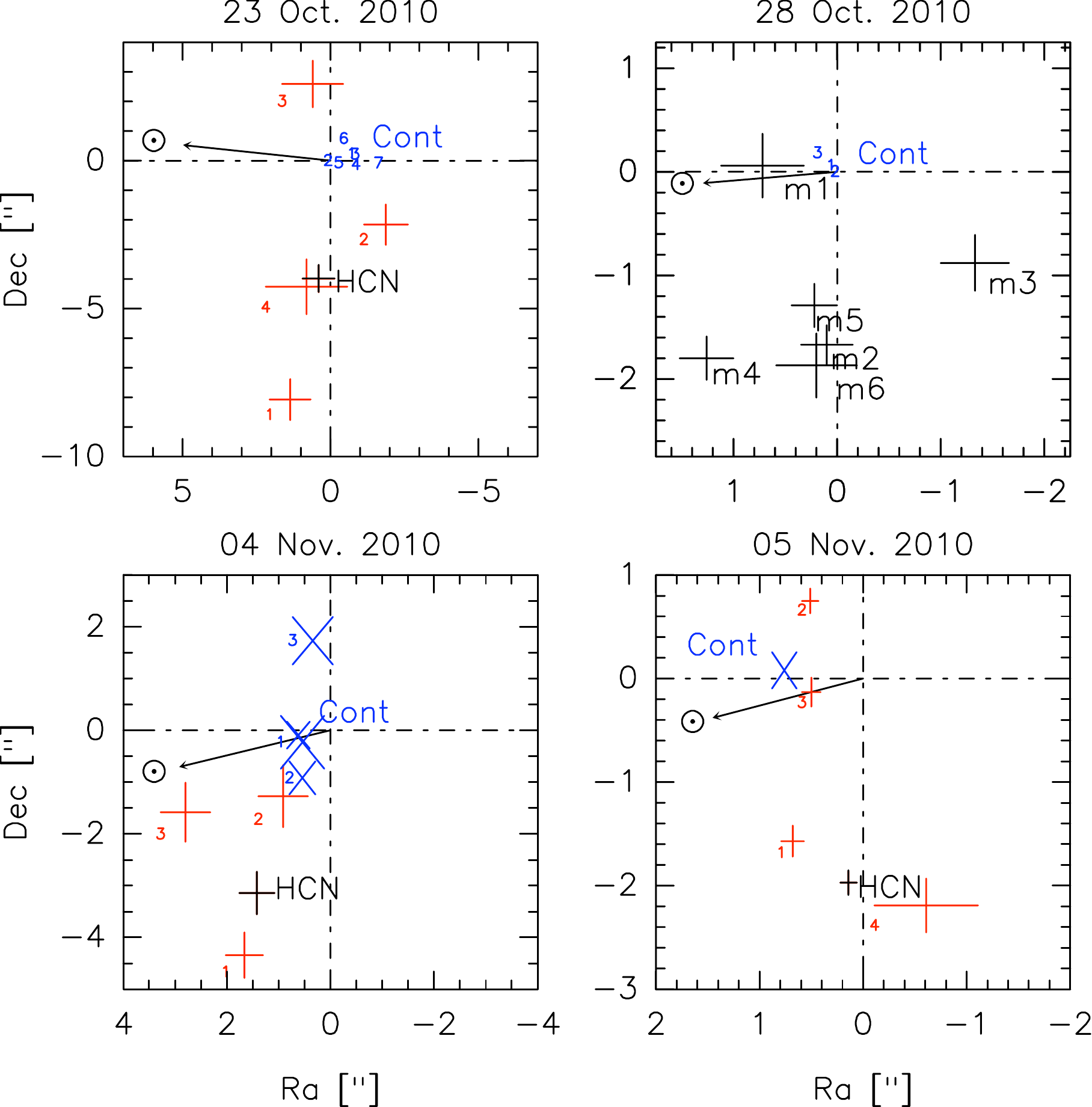}
 \caption{Positions of the peak brightness of HCN and CH$_3$OH lines and of the continuum emission (blue symbol) with respect to the nucleus position given by the ephemeris at (0,0) (Table~\ref{tab-obs}).  Positions were determined by fitting a point source to the interferometric data. 
For molecular lines, positions from daily-average data are shown by black crosses. 
The successive positions along the observing day are displayed by red crosses and numbers for the HCN lines,  and blue numbers and/or crosses for the continuum. In the upper right frame, the labels m1 to m6 refer to the 6 observed methanol lines. The sizes of the crosses correspond to 1-$\sigma$. The arrow indicates the direction of the Sun.
}
 \label{fig-pos}
 \end{center}
 \end{figure}

\subsubsection{Positions of the peak brightness}
\label{sec:pos}

Figure~\ref{fig-pos} shows the positions of the maximum line and continuum emission determined from the maps.
While the continuum peak almost coincides with the latest ephemeris solution, all HCN and CH$_3$OH peak positions derived from daily-averaged data are shifted South-East by 2$^{\prime\prime}$ to 4$^{\prime\prime}$. The largest offsets are observed at 3.4~mm wavelength, i.e., when the interferometric beam is the largest. This trend is consistent with a displacement related to  
asymmetries in the brightness distribution. The time variations of the HCN brightness peak position are also reported in Fig.~\ref{fig-pos}. The successive positions are not similar and move approximately along the North-South (NS) axis, whereas the Sun direction was at P.A. = 83$^\circ$ to 105$^\circ$ from 23 October to 5 November. CN optical images obtained from 27 October to 8 November 2010  \citep{lar+11,sam+11,knisch11,Wan+12} show two almost spatially opposite jet features, northward and southward, with the relative strengths of the two jets varying with time. There are therefore resemblances in HCN and CN coma morphologies.          
In addition, the variability detected in CN correlates well with the cyclic changes in HCN during the active phases \citep{dra+12,Wan+12}.
The expected dust counterparts of the CN jets are not detected in optical images \citep{lar+11,muel+13}.

\section{Sources of HCN from temporal variations}
\label{sec-jet}

\begin{figure}
  \begin{center}
\includegraphics[width=6cm]{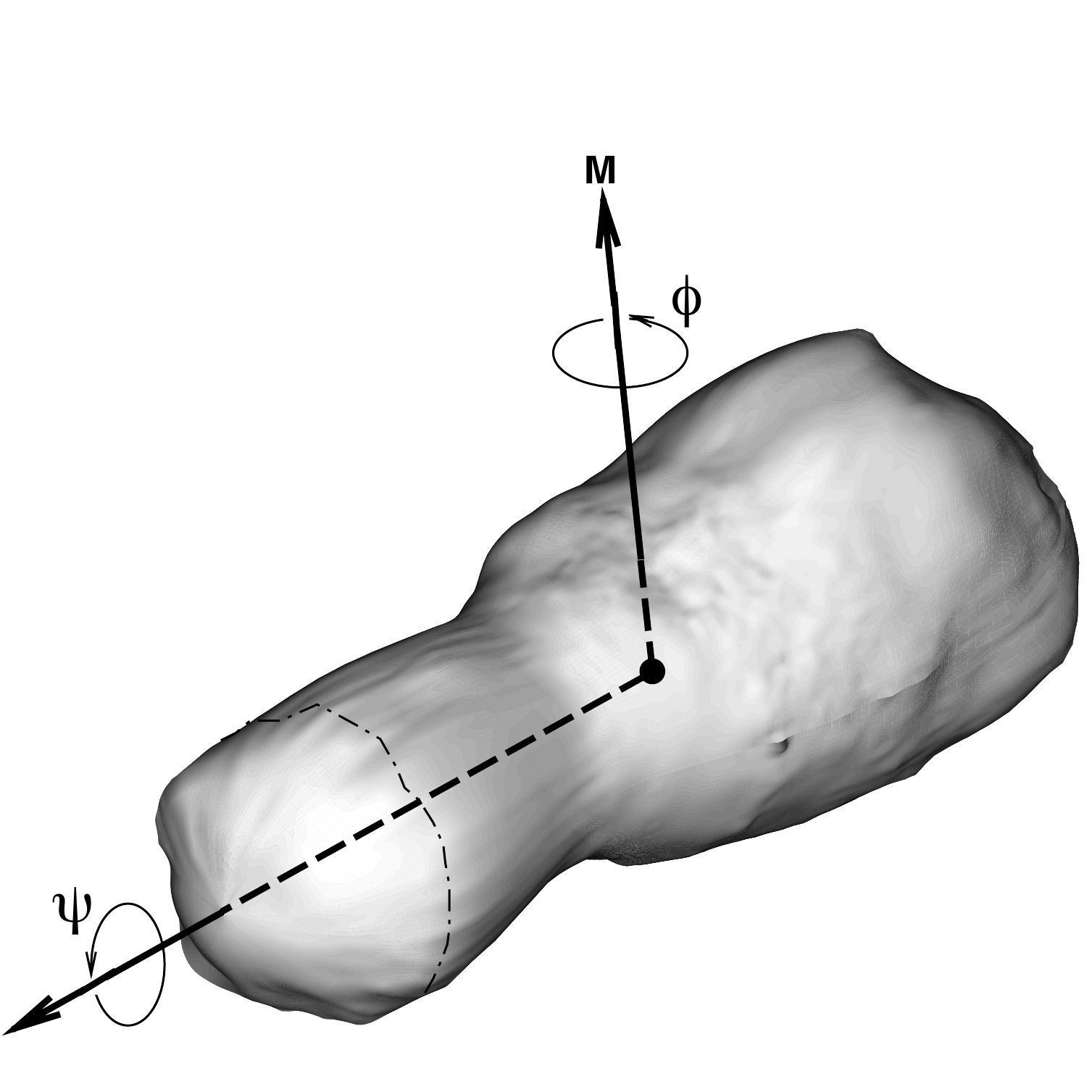}
\caption{Shape model of the nucleus of 103P/Hartley 2 from \citet{thomas+12}. The
direction of the total angular momentum {\bf $M$} and the long axis are
indicated. The nucleus rotates in the long-axis mode, with $P_{\phi}$
$\sim$ 18.4 h and $P_{\psi}$ $\sim$ 27 h at the
time of the EPOXI flyby \citep{belton+12}.
The dashed-dotted line corresponds to the latitude of 70\tdeg.
}
 \label{fig-nuc}
 \end{center}
\end{figure}

\subsection{Context and approach}
\label{sec:context}

Images from the EPOXI mission show that comet 103P/Hartley~2 is a bi-lobed, elongated, nearly axially symmetric comet \citep[Fig.~\ref{fig-nuc},][]{ahe+11,thomas+12}.
EPOXI revealed strong asymmetries in the spatial distribution of dust and gaseous species. The small lobe (hereafter also refered as to SL) of the nucleus was found more productive in CO$_2$, dust and icy grains than other regions,   whereas enhanced water density was observed above the waist \citep{ahe+11}. 
 \citet{dra+11} showed that the outgassing pattern of HCN repeated best every
three 18.3-hr rotation cycles, denoted as A, B, and C, with the middle of the system
(phase 0.5 of Cycle B) defined at the moment of the EPOXI flyby. They interpreted
this behavior as caused by excitation of the nucleus rotation state, later fully
quantified by  \citet{belton+12}. This three-cycle periodicity is also observed in the light curves of CO$_2$, H$_2$O and dust observed by EPOXI \citep{ahe+11,besse+12}, and for the morphology of the CN coma \citep{knisch11,Wan+12}.
The three cycles A, B, and C covered by the 3 to 6 November date range are indicated in Fig.~\ref{fig-Qcurve1}. 

 Using the line profiles of HCN and CH$_3$OH, \citet{dra+12} identified three distinct components: two jets producing, respectively,  the redshifted and blueshifted peaks (Sect.~\ref{line_I_V}), 
and an isotropic component producing symmetric double-peak line profiles, associated with sublimating icy grains. The blueshifted jet was tentatively 
associated with the CO$_2$ jet emanating from the small lobe, whereas the redshifted jet with
the area producing the H$_2$O density enhancement above the waist, and supposedly close to the region of polar night, in order to explain its short duration.

From the knowledge of the spin state, its orientation in space and kinematics, it is possible to 
compute the insolation of surface elements as the nucleus rotates  \citep[this was also done by][]{belton+12}. If we further assume that 
the outflow occurs perpendiculary to the surface, the direction of the outflow with respect 
to the line of sight (so-called aspect angle $Ap$, $Ap$ = 0 for a vector directed towards Earth) can be computed and compared with the observed velocity shifts. We have investigated how the time evolutions of the insolation and $Ap$ distribution compare with the evolution of the HCN production rate and line velocity shift, respectively, in order to constrain the sources of HCN. 
 In contrast to HCN, the emission properties of the methanol lines strongly depend on the coma temperature  
\citep[see Fig.~7 of][]{dra+12}. Hence, CH$_3$OH line shapes and intensities are strongly affected by temporal and spatial variations of the gas temperature in 103P's coma (see Sect.~\ref{sec-temp}), so that a similar study for CH$_3$OH would require to consider these effects. This is beyond the scope of this paper.

We used the spin parameters provided by \citet{belton+12} to compute, as a function of time, the cosine of the angle of the normal to the surface elements with the Sun direction, and with the Earth direction.                                   
According to \citet{belton+12}, at EPOXI closest approach, the long axis was precessing with a period of $P_{\phi}$ $\sim$ 18.4 h around the rotational angular momentum  {\bf $M$} (RA, Dec; J2000 = 8$^{\circ}$, 54$^{\circ}$), and tilted to  {\bf $M$} by 81.2$^{\circ}$ (Fig.~\ref{fig-nuc}). In addition, the body was rolling around the long axis with a period of $P_{\psi}$ $\sim$ 27~h. The shape model of \citet{thomas+12} was used and we adopted the geographic coordinate system ($lat$, $long$)  defined by \citet{thomas+12}, with the tip of the SL defining the north pole (Fig.~\ref{fig-nuc}). 
For simplicity, we have neglected shadowing effects when computing the insolation of surface areas. 

\subsection{Comparison of HCN temporal variations with nucleus illumination}
\label{sec:5.2}

\begin{figure}
 \begin{center}
\includegraphics[angle=270,width=9cm, bb = 100 50 537 650]{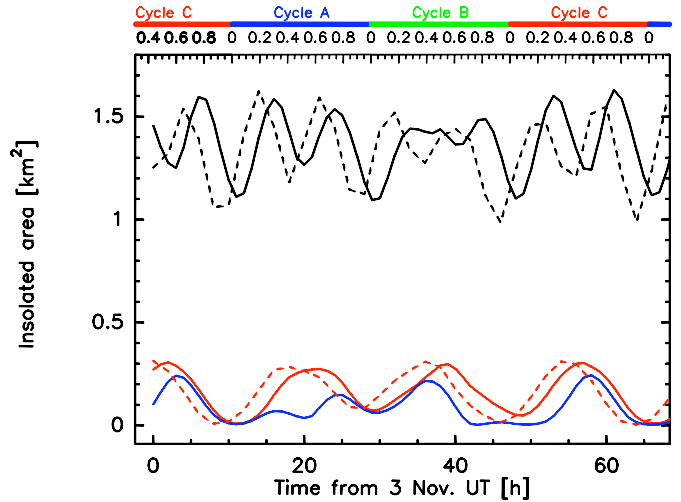}
 \caption{Insolated cross-section of 103P's nucleus as a function of time. Results obtained with the rotational angular momentum {\bf $M$} directed towards (RA, Dec; J2000 = 8$^{\circ}$, 54$^{\circ}$) and (RA, Dec; J2000 = 83$^{\circ}$, +53$^{\circ}$) \citep{belton+12} are in plain and dashed lines, respectively. The black curves correspond to the whole nucleus. The curves in red consider areas at $lat$ $>$ 70$^{\circ}$ and $long$ between 0--360$^{\circ}$ (small lobe). The curve in blue color consider regions on the small lobe with $long$ = 0 to 180$^{\circ}$.}
 \label{fig-section}
 \end{center}
 \end{figure}

\begin{figure}
%\begin{center}
%\includegraphics[angle=270,width=12cm]{./map-illu-select.eps}
\hspace{-0.2cm}\includegraphics[angle=270,width=13cm]{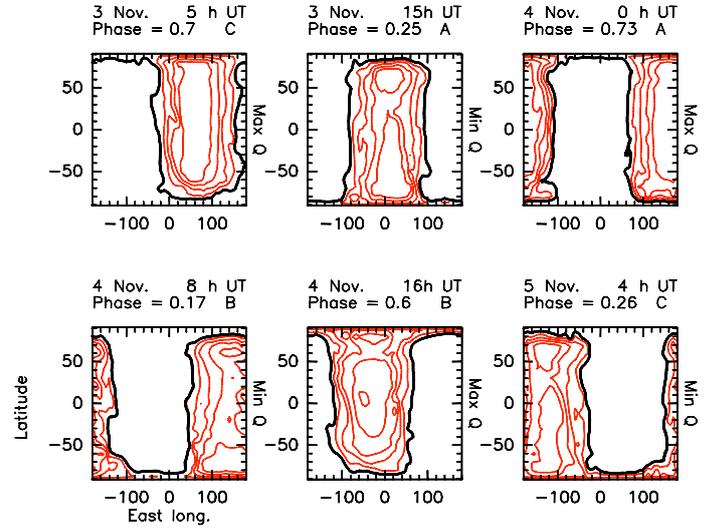}
 \caption{Illumination of the nucleus of 103P/Hartley 2 at times corresponding 
 to minimum or maximum HCN production, as indicated in the right part of the plots.
 Isocontours correspond to the cosine of the solar zenith angle, shown by step of 0.2 (illuminated regions). The contour in black color is the terminator.
}
 \label{fig-illu}
% \end{center}
 \end{figure}

Figure~\ref{fig-section} shows the variation of the illuminated surface of the nucleus (surface elements are weighted by the cosine of the solar zenith distance) for the date range considered in Fig.~\ref{fig-Qcurve1}. It follows a complex pattern 
related to the non-principal-axis rotation mode and elongated shape of the nucleus. The illuminated fraction of the small lobe \citep[defined as $lat$ $>$ 70$^{\circ}$,][]{thomas+12} is also plotted in the figure. The  measured HCN production
rate varied by a factor of $\sim$ 4 from minimum to maximum activity (Fig.~\ref{fig-Qcurve1}).  
Minimum HCN activity is observed on 3 November $\sim$ 15 h (Cycle A, phase 0.25), 4 November 8 h (Cycle B, phase 0.17), and 5 November 4 h UT (Cycle C, phase 0.26). It is striking that these times correspond approximatively to the times when the surface of the SL exposed to the Sun is minimum (Fig.~\ref{fig-section}, the time lag is discussed at the end of this section). Figure~\ref{fig-illu} shows maps of the solar illumination of the 
nucleus at the times of observed minimum and maximum HCN production. Maximum HCN production is observed when the tip of the SL is into sunlight, whereas the tip of the SL is in shadow at the times of minimum production. Interestingly, maps for 3 November, 5 h UT, and 5 November 4 h UT, both in Cycle C, corresponds to maximum and minimum HCN production, respectively, whereas they are complementary regarding illumination (rotational phases differ by 0.44), the nucleus regions in sunlight for the former time being in shadow for the latter time, and vice-versa.  This leads us to the conclusion that the evolution of the HCN production rate was mainly modulated by the insolation of the SL. 
 This conclusion also applies to CH$_3$OH, the production curve of which is similar to that of HCN, albeit generates smaller amplitude variations \citep{dra+12}.
We can also conclude that the surface areas of the SL were active mainly when in sunlight.

\begin{figure}
%\begin{center}
%\includegraphics[angle=0,width=10cm]{./map-illu-aspect-select-dv-1.eps}
\hspace{-1cm}\includegraphics[angle=0,width=12cm]{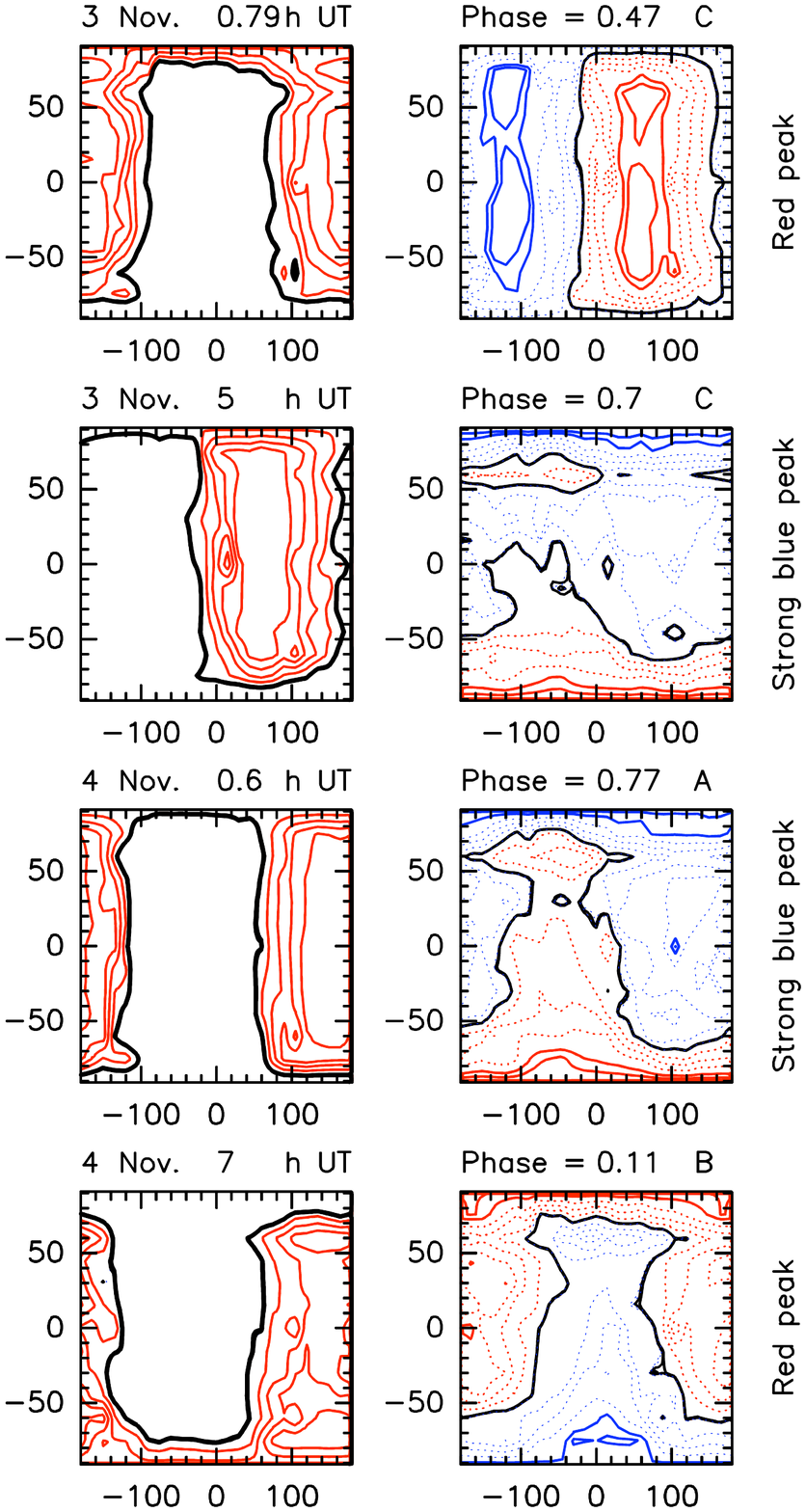}
\vspace{-2cm}
\includegraphics[angle=0,width=10cm, bb = 51 300 567 710]{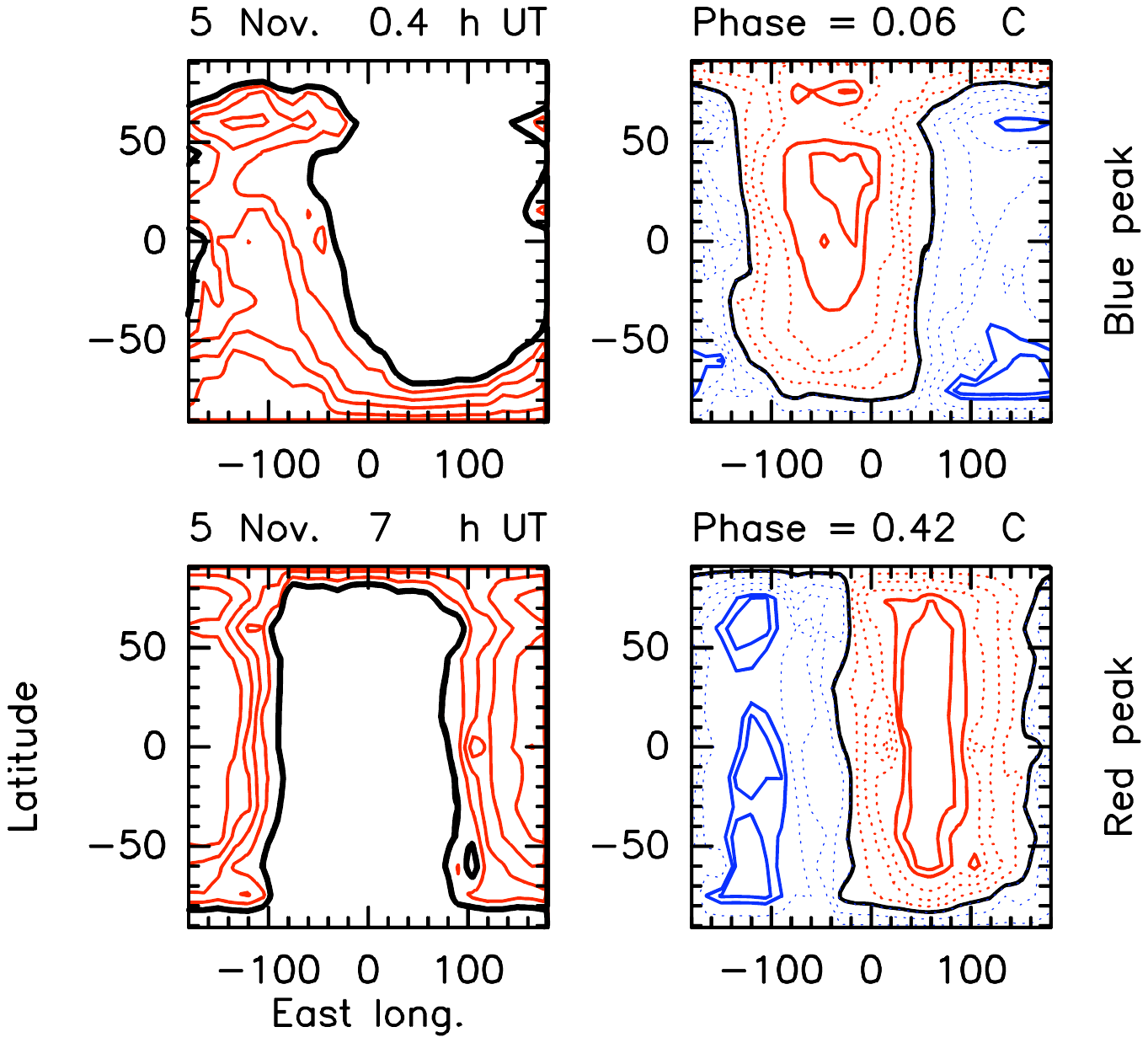}

 \caption{Illumination of the nucleus of 103P/Hartley 2 (left) and maps of equi-$Ap$ (right) at times when HCN spectra showed a strong blue peak or enhanced signal at positive velocities, as indicated in the right part of the plots. Contours and color coding for illumination maps are as for Fig.~\ref{fig-illu}. 
In maps of equi-$Ap$, isocontours correspond to the cosine of $Ap$, shown by step of 0.2 for 
$\vert$cos($Ap$)$\vert$ $<$ 0.8 (dashed contours), and by step of 0.1 for $\vert$cos($Ap$)$\vert$ $>$ 0.8 (solid contours). Normal to surface areas projecting towards and outwards the Earth direction are in blue and red, respectively, and are delimited by the black isocontour (cos($Ap$) = 0).}
 \label{fig-illu-aspect}
%\end{center}
 \end{figure}

We show in Fig.~\ref{fig-illu-aspect} (left) maps of the solar illumination at the times when the HCN lines displayed either  their strongest blueshift, or a well-defined redshift peak. For each of these times, we also show how the normal of the surface elements projects onto the line of sight, by plotting isocontours of  cos($Ap$) (Fig.~\ref{fig-illu-aspect}, right). The strong blueshifts  on 3 November 5 h UT (Cycle C, phase 0.7)
and 4 November 0.6 h UT (Cycle A, phase 0.77) are observed at times where the SL is illuminated and facing the Earth (cos($A_p$) reaching values $>$ 0.8). Hence, the blueshifted peak observed at these times in the HCN spectra, is associated with material produced from the SL. The source region of the blue peak observed on 5 November 0.4 h UT (Cycle C, phase 0.06) at time of minimum activity is
less straightforward. Regions at $lat$ $>$ 75$^{\circ}$ are excluded as they were in shadow and were facing anti-Earthward.
 This blue peak might be related to jets originating from SL latitudes $\sim$ 60--75 $^{\circ}$ with longitudes  180--220$^{\circ}$ E, as these regions were in sunlight and were pointing Earthward and  South (i.e., according to the observed displacement of the HCN peak brightness, Fig.~\ref{fig-pos}) at times when the icy grains producing HCN were released (see discussion later on). As seen in Fig.~\ref{fig-Qcurve1}, outgassing from illuminated SL areas (most SL areas being in darkness) satisfactorily  explains both the small production
rate and large HCN line velocity shift measured at the beginning of Cycle C. Alternatively, outgassing from 
other regions, including the waist and mid latitudes on the large lobe, could also explain the observations.

\citet{dra+12} argue that the single redshifted peak originated from a different region than the single blueshifted peak, 
associated to SL outgassing. However, since the SL is much more active than other regions, it may also be responsible for a redshifted peak. Indeed, the development of the redshifted peak coincides with the disappearance of the blueshifted peak (see spectra in Fig. 13 of \citet{dra+12}), which is possibly a projection effect, since the mean outflow from the SL changes its direction with respect to the Earth, as the nucleus rotates. For exemple, CO $J$(2--1) spectra of comet C/1995 O1 (Hale-Bopp) acquired over one rotation period near perihelion displayed enhanced emission moving from positive to negative Doppler velocities in the nucleus velocity rest frame, due to the presence of a CO rotating jet active day and night \citep{boc+09hb}. To produce enhanced signal at red velocities, the surface regions on the SL need to be illuminated at times when their normal is pointing in the half-space opposite to the Earth direction (assuming that the activity proceeds only during day time for 103P). These conditions are fulfilled for Cycle C, phase = 0.42--0.47 (3 Nov. 0.79 h UT and 5 Nov. 7 h UT) (Fig.~\ref{fig-illu-aspect}).  However, the red peak observed at phases 0.9--1 of Cycle A to 0.2 of Cycle B can be attributed to outgassing 
from illuminated SL mid-latitudes ($<$ 70$^{\circ}$) or to remnant activity from SL high latitudes into darkness (Fig.~\ref{fig-illu-aspect}). Contribution from other regions (e.g., mid latitudes on the large lobe or the waist) is possible, though regions expected to produce a well-defined redshift peak are not (their local time corresponds to evening) or poorly illuminated. Interestingly, both this red peak and the blue peak observed at beginning of Cycle C can be explained by outgassing from the SL at 60--70$^{\circ}$ latitudes located above the waist. Alternatively, 103P/Hartley 2 possibly displayed night-side jet activity, but this conclusion relies on the assumption that the spin parameters 
that we are using are fully accurate (see discussion in Sect.~\ref{sec:5.4}).
   We refer the reader to \citet{dra+12} who argue that the source region of the redshifted peak should be located on the waist, to explain both HCN and CH$_3$OH data line profiles and their time evolution.

We show in Fig.~\ref{fig-illu-aspect-CA} the illumination and $A_p$ maps at time of the EPOXI closest approach (Cycle B, phase of 0.5). The whole longitude range of the SL was illuminated and the long axis was approximately in the plane of the sky (aspect angle of 75$^{\circ}$). Outgassing from $long$ = 0--180$^{\circ}$ E would produce a blueshifted peak, with the rest producing a redshifted peak. This might explain qualitatively the double-peak structure of the HCN profile observed in Cycle B, phase = 0.5 (see the HCN spectrum on 2 November $\sim$ 7 h UT in Fig. 13 of \citet{dra+12}),  though the detailed interpretation of the whole HCN production curve suggests that SL regions at $long$ = 180--360 $^{\circ}$ E were less active (Sect.~\ref{sec:5.4}). Finally, worth is mentionning that spectra with a double-peak structure are expected in a number of circonstances which apply here, i.e., a well resolved coma, production from distributed sources \citep{dra+12}, and spiralling jets \citep{boc+09hb}.

\subsection{Comparison of HCN, CH$_3$OH, H$_2$O and CO$_2$ production curves}
\label{sec:5.3}     
We have shown that gas production curves are highly modulated by the insolation of the SL. 
However, there is not a perfect correspondence between the times of maximum HCN production and maximum SL illumination. For exemple, on 3 November  
maximum HCN production was observed at $\sim$ 6 UT (Cycle C, phase = 0.77), whereas maximum illumination of the SL occurred at 2 h UT (phase = 0.54). There 
is a similar time shift between minimum activity and minimum SL illumination. The time lag is long with respect to the expected ``beam delay'' of $\sim$ 0.2 h \citep[see][where we scaled computed instrument responses to production rate variations using the beam size and geocentric distance appropriate to the 103P observations]{biv+07}. We compared the production curve of HCN to those of H$_2$O and CO$_2$ recorded by EPOXI from 7 to 9 November \citep{besse+12}. For CO$_2$, maximum production is observed  
at times corresponding to phases 0.62 of Cycle C, 0.74 of Cycle A, 0.44 of Cycle B. For H$_2$O, the peaks of production are at phases 0.69 of Cycle C, 0.87 of Cycle A, and phase 0.52 of Cycle B, i.e., delayed by 1.44 h (peaks in phase B and C) and 2.4 h (peak of cycle A) with respect to the CO$_2$ production curve.  For comparison, maximum HCN production is at phases $\sim$ 0.77 (C) 
(0.70 (C) from a fit of the overall evolution in the phase range 0.4--0.9), 0.82 (A), and $\geq$ 0.5--0.6 (B). Hence, the H$_2$O and HCN production curves are in average in phase, suggesting that these molecules have common sources. A perfect phasing between CH$_3$OH and HCN production is also observed \citep{dra+12}. This time lag with the CO$_2$ production curve can be related to the production of H$_2$O, CH$_3$OH, and HCN from subliming icy grains, whereas CO$_2$ molecules were released from the nucleus. Significant contribution of grain sublimation to the production of volatiles is supported by numerous measurements  \citep[e.g.,][]{ahe+11,mum+11,del+11,dra+12,kni+12}.

\subsection{Detailed analysis of the HCN production curve} 
\label{sec:5.4}
  
\begin{figure}
 \begin{center}
\includegraphics[angle=0,width=11cm]{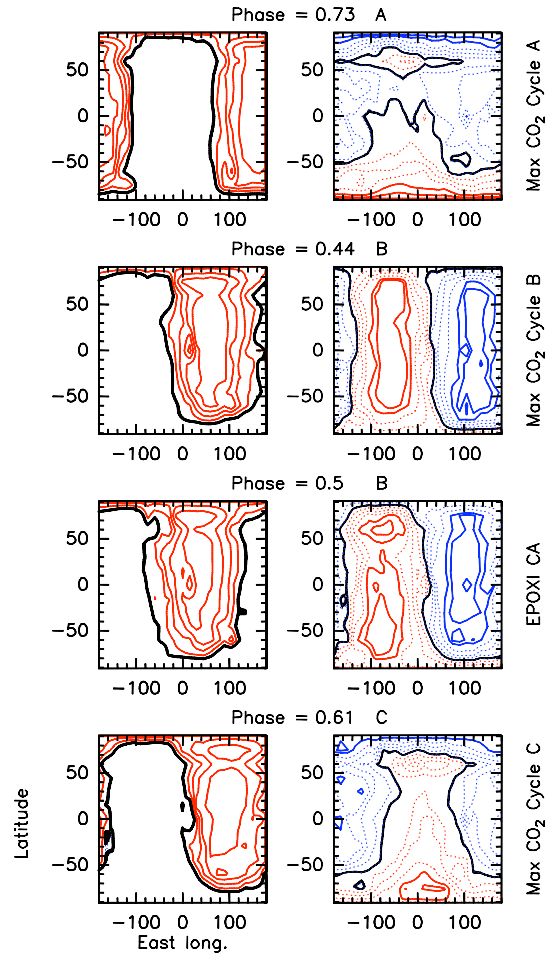}
 \caption{Illumination of the nucleus of 103P/Hartley 2 (left) and maps of equi-$Ap$ (right) at 
 times of maximum CO$_2$ production for cycles A, B, and C, and at time of EPOXI closest approach (4 November 14 h UT). See Fig.~\ref{fig-illu-aspect}.}
 \label{fig-illu-aspect-CA}
 \end{center}
 \end{figure}

We now compare the CO$_2$ production curve to the variation of the illumination of the SL area.  
There is a small time difference between SL maximum insolation and the peaks in the CO$_2$ production curve, which varies from cycle to cycle: $\sim$ --2.1 h, +2 h, and --1.4 h for Cycles A, B and C respectively.
A possible explanation for this time lag is that the non-principal axis rotation parameters provided by \citet{belton+12} provide only an approximate description of the spin state of 103P/Hartley 2. Indeed, there is still no agreement on the direction of the rotational angular momentum {\bf $M$} \citep[see][and references therein]{belton+12}, whereas the time variation of SL insolation is very sensitive to the space direction of {\bf $M$}. For exemple, using instead the first solution provided by \citet{belton+12} (RA, Dec; J2000 = 83$^{\circ}$, +53$^{\circ}$), which is based on the directions of the long and intermediate principal axes of \citet{thomas+12}, the discrepancy is worse (Fig.~\ref{fig-section}). The time lag is more likely due to local variations of the outgassing on the SL, related, e.g., to a non uniform distribution of active areas. 
This scenario is supported by the different strengths $S$ of the peaks at each cycle \citep{besse+12} ($S$(C) $>$ $S$(B) $>$ $S$(A) for H$_2$O and CO$_2$), and by the difference of spacing between the peaks (CO$_2$: 13 h, 21.6 h and 20.6 h, for A--B, B--C, and C--A, respectively). Indeed, whereas the overall insolation of the SL shows maxima of the same strength for each cycles (Fig.~\ref{fig-section}), this is not expected for specific regions on the SL due to the rolling along the long axis. Simulations show that, for specific regions of the SL defined by their longitude range, the spacing between the peaks are typically 11/21/22 h, i.e., at a frequency which corresponds to (1/$P_{\phi}$+1/$P_{\psi}$), and half this value. Figure~\ref{fig-section} shows the variation of the insolation of SL area at longitudes between 0--180$^{\circ}$. The times of peak of production of CO$_2$ are correctly reproduced (within 20 to 50 min). The relative strengths of the peaks are qualitatively explained, in particular $S$(C) $>$ $S$(A). However, the observed $S$(C)/$S$(B) ratio is higher than the ratio of the peaks in the simulation, the latter being only slightly above unity because, on the basis of the rotation parameters of \citet{belton+12}, almost the same areas of the nucleus were exposed to the Sun at the times of peak activity during cycles B and C (Fig.~\ref{fig-illu-aspect-CA}). 

Figure~\ref{fig-Qcurve1}a--b shows the time evolution of the insolation and insolation-weighted cos($Ap$) of the $long$ = 0--180$^{\circ}$ SL region for a comparison with the measured HCN production rate and line velocity shift. We shifted the curve of the insolation by +1.71 h (which corresponds to the mean delay between H$_2$O and CO$_2$ peaks of production), to account for production of HCN by grains. The agreement with the production curve is overall satisfactory, taking into account that outgassing from regions other than the SL is not considered.

An overall good agreement is also obtained between the shape of the insolation-weighted cos($Ap$) curve and the HCN line velocity shift, when not applying a +1.71 h time shift to the model (Fig.~\ref{fig-Qcurve1}b). Indeed, though HCN molecules are presumably produced from grains, their large velocity ($\sim$ 0.6--0.7 km s$^{-1}$ based on the line width and the Doppler velocity of the blueshifted peak) may indicate that they share the velocity field of the ambient gases through collisional interactions  (production by small grains with gas-like velocity is not favored, Sect.~\ref{sec-radext}). 
Whether the regions where HCN gassesgre
 are produced are dense enough to change the isotropic velocity of the molecules released from the icy grains might be questionable. In Sect.~\ref{sec-radext}, we show that the interferometric maps suggest a typical scale length of HCN production of 500--1000 km. According to the model of \citet{fougere+13}, at these distances above the active SL, collisions are still important though possibly not numerous enough for a true fluid-equilibrium regime (Fougere, personal communication, found at 500 km a Knudsen number $K_n$ = 0.22 for the density gradient, and $K_n$ = 0.50 for the temperature gradient, whereas free-molecular flow requires $K_n >> 1$, and the fluid regime requires $K_n < 0.05$ ). Therefore, the question remains opened. Note that, assuming that the HCN molecules follow the ambient velocity field related to the geometry of production, a small time shift of 0.2--0.4 h in the cos($Ap$) curve, not applied in the model curve shown in Fig.~\ref{fig-Qcurve1}b, should have been in principle considered. We observe in Fig.~\ref{fig-Qcurve1}b a discrepancy for $\Delta v$ at phase 0.1 of Cycle A, which is not critical for our interpretation since at that time the SL was in shadow and almost inactive. The observed moderate blueshift is consistent with some outgassing from the large lobe of the nucleus which was into sunlight and pointing the Earth at that time. The discrepancy for phase 0.4--0.6 of Cycle B shows also that regions other than the SL  $long$ = 0--180$^{\circ}$ region were possibly contributing to HCN production (see discussion in Sect~\ref{sec:5.2}).

We compare in Fig.~\ref{fig-Qcurve1}c the direction of the offset of the HCN peak brightness 
measured in the HCN maps with the mean direction of the outgassing from insolated area at $long$ = 0--180$^{\circ}$ on the SL. There is a good agreement with the data from 4 November (phase 0.9 of cycle A to 0.1 of cycle B). However the rapid variation of the position offset observed on 5 November, at a time when most of the SL
was in shadow, is not reproduced by the model, showing that other localized sources of enhanced activity need to be considered to interpret the data.  These sources are possibly responsible for the second CN jet observed in optical images (Sect.~\ref{sec:pos}).

We used geometric considerations to interpret the gaseous production curves. An extensive modelling is beyond the scope of this paper. However it is worth mentionning that we did not consider a possible time lag related to the production of CO$_2$ from below the surface. For a thermal inertia in the range 50 (250) MKS \citep{grou+12}, the subsurface temperature at depths of 2 (10) cm is shifted by 3 h in the afternoon. The good correlation between SL illumination and CO$_2$ production suggests that CO$_2$ is originating from small depths.

\section{Radial distribution of HCN}

\label{sec-radext}

As discussed in Sect.~\ref{sec-jet}, the HCN production curve  suggests a large contribution from sublimating icy grains. Icy grains were also likely an important source of CH$_3$OH. Indeed the variability of the CH$_3$OH production rate correlates well with that of HCN \citep{dra+12}. We investigated whether there is evidence for extended production of molecules in the Plateau de Bure data, by comparing the 
interferometric  ($F_{\rm Int}$) and single-dish ($F_{\rm SD}$) fluxes, following \cite{boi+07} and \cite{boc+10}. An extended source production, if resolved out by the interferometric beam, 
would result in a ratio $R = F_{\rm SD}/F_{\rm Int}$ higher than the value expected
for a pure nuclear production. On the other hand, this study requires to account for the optical thickness of the lines in model calculations. Indeed, since opacity effects are more important for smaller beams, the $F_{\rm SD}/F_{\rm Int}$ intensity ratio is higher for optically thick conditions. 

We present in Table~\ref{tab-ratios} line intensity ratios $R_{\rm obs}$ that can be reliably used for this study.
For methanol observed on 28 October, the six emission lines were combined to increase the signal-to-noise ratio and the whole observing period was considered. For HCN observed on 23 
October and 4 November, we selected time ranges where the single-dish data displayed limited temporal variations. For HCN on 5 November, the full data set was again considered. Time ranges for single dish and interferometric observations (see Table~\ref{tab-ratios}) were selected so that approximately the same molecules contribute to the fluxes measured in the two modes (see Sect.~\ref{line_I_V}).

We used the excitation and radiative transfer model described in Section~\ref{sec-temp} to compute 
synthetic On-Off and interferometric fluxes considering a nuclear 
or extended production of the molecules \citep[see also][]{boi+07}. This
model takes into account opacity effects, which affect significantly the interferometric fluxes.
The cross-section for HCN--H$_2$O collisions was taken equal to  
$\sigma_c = 1\times 10^{-14}$ cm$^2$ \citep{biv+99}. 
Extended distributions were investigated using the Haser formula
for daughter species with the parent and daughter scale
lengths $L_{\rm p}$ and $L_{\rm d}$ as parameters. Parent scale lengths up to 1000 km
were considered. $L_{\rm d}$ corresponds to the photodissociative scale length, which does not affect the calculations as the field of view is probing the inner coma. We considered three cases for the production sources of the molecules: nuclear ($Q^{\rm N}:Q^{\rm E}$ = ($100:0$)), extended ($Q^{\rm N}:Q^{\rm E}$ = ($0:100$)), 
and composite ($Q^{\rm N}:Q^{\rm E}$ = ($50:50$)), where $Q^{\rm N}$ and $Q^{\rm E}$ are the 
production rates from the nucleus and extended source, respectively. 

Finally, calculations showed that model results are sensitive to the assumed azimuthal distribution of the gas. We therefore focussed on the HCN $J$(1--0) data acquired on 4 November from 1 to 3.5 h UT (Table~\ref{tab-ratios}), for which the outgassing pattern is well understood (Sect.~\ref{sec-jet}). According to the model presented in Sect.~\ref{sec-jet}, the mean orientation of the HCN vent from the small lobe was at an aspect angle of 65$^{\circ}$ with respect to Earth direction. To account for the observed Doppler shift of --0.2 km s$^{-1}$ (Fig.~\ref{fig-Qcurve1}), we set the fraction of HCN molecules in the vent to a value $j$ = 75$\%$.  We assumed that $Q^{\rm N}:Q^{\rm E}$ is the same in the vent and outside the vent. Whereas we present model calculations for a constant kinetic temperature in the coma of 40 K, similar results and conclusions are obtained when considering a temperature gradient, or a larger temperature in the vent (60 K) than in the rest of the coma (30 K) (Sect.~\ref{sec-temp}). 

\begin{figure}
 \begin{center}
\includegraphics[width=7.8cm]{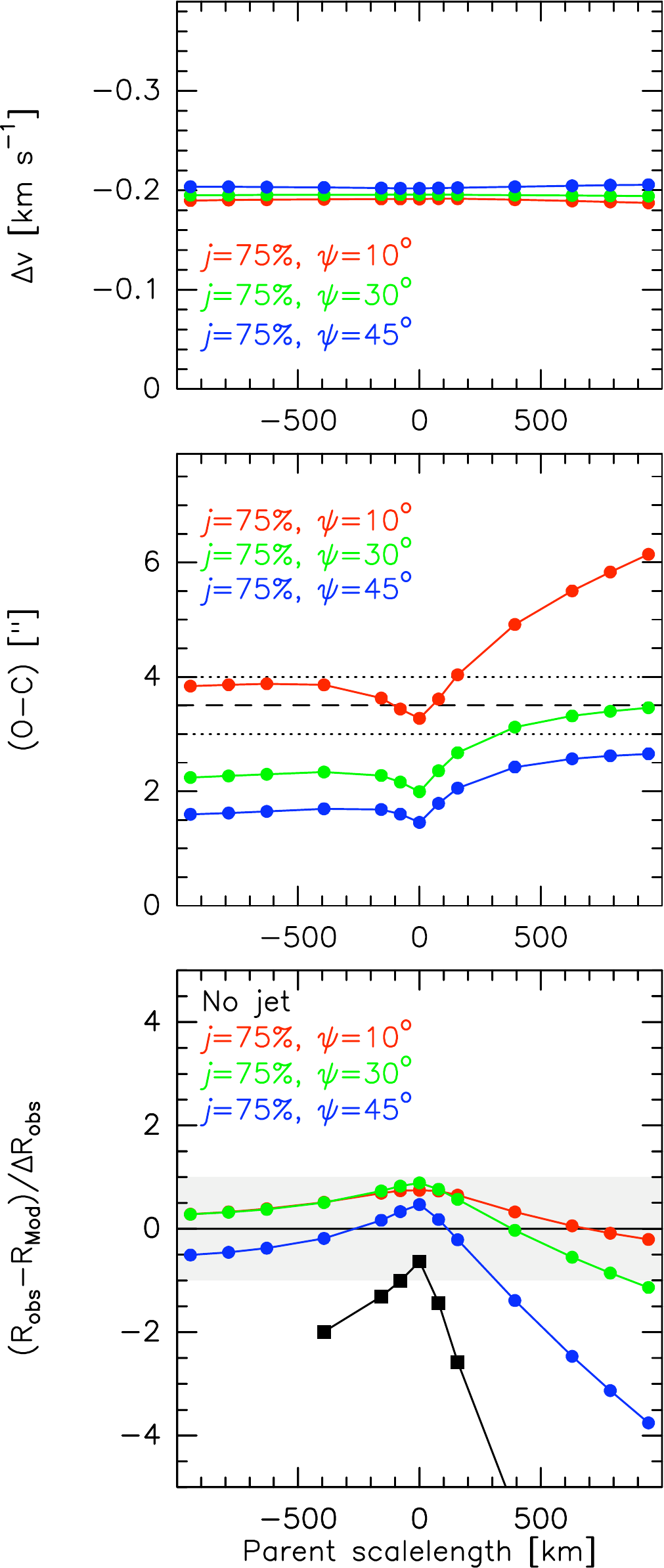}
 \caption{Model simulations for the $J$(1--0) HCN observations undertaken on 4 November (1--3.5 h UT) with the Plateau de Bure interferometer. The modelling considers an asymmetric HCN coma where a fraction $j$ = 75 $\%$ of the HCN molecules are contained in a jet of semi-aperture $\psi$ = 10\tdeg~(red symbols), $\psi$ = 30\tdeg~(green symbols), and  $\psi$ = 45\tdeg~(blue symbols). The angle between the jet and the Earth direction is 65\tdeg. Results for an isotropic coma are shown with black symbols.
{\bf Top:} Line velocity offset in single-dish mode, consistent with the observed value of --0.2 km s$^{-1}$. {\bf Middle:} Offset of the brightness peak of the synthesized map with respect to the nucleus position. 
The dashed line represents the observed offset, the dotted lines define the $\pm 1 \sigma$ values.
{\bf Bottom: } $\chi$ = $(R_{\rm Obs}-R_{\rm Mod})/\Delta R_{\rm Obs}$; the grey region corresponds to --1 $<$ $\chi$ $<$ 1. Results plotted in the right part of the plots (positive scale lengths) are for pure extended distributions (i.e., ($Q^{\rm N} : Q^{\rm E}) = (0 : 100)$). Results for composite distributions $(Q^N : Q^E) = (50 : 50)$
 are shown in the left part of the plot (negative scale lengths). }
 \label{fig:HCN-rad}
 \end{center}
\end{figure}

Figure~\ref{fig:HCN-rad} shows the model results for a HCN jet with semi-aperture angles $\psi$ of 10, 30 and 45$^{\circ}$, together with results obtained for an isotropic HCN distribution. This figure displays: i) the velocity shift of the HCN $J$(1--0) line in single-dish mode; ii) the position offset of the peak intensity with respect to the nucleus position; iii) the quantity     
$\chi$ =  $(R_{\rm Obs}-R_{\rm Mod})/\Delta R_{\rm Obs}$, where $R_{\rm Mod}$ is the intensity ratio computed by the model and $\Delta R_{\rm Obs}$ is the uncertainty on $R_{\rm Obs}$. These three quantities are plotted as a function of the parent scale length $L_p$ of HCN, with negative parent scale lengths refering to the composite case ($Q^{\rm N}:Q^{\rm E}$ = ($50:50$)), positive parent scale lengths refering to the extended case ($Q^{\rm N}:Q^{\rm E}$ = ($0:100$)), and $L_{\rm p}$ = 0 corresponds to nuclear production. The third quantity allows us to estimate the goodness of the fit for the intensity ratio $R_{\rm Obs}$ = 19.1 $\pm$ 7.0 measured on 4 November (Table ~\ref{tab-ratios}); values of this quantity between -1 and 1 are considered as satisfactory. The total HCN production rate ($Q^{\rm N} + Q^{\rm E}$) was taken equal to 2.0 $\times 10^{25}$~s$^{-1}$, which corresponds to the value which reproduces the line area observed in single-dish mode when a pure nuclear 
 source is assumed.  

The results for an isotropic distribution of HCN shows that this assumption could lead to erroneous conclusions concerning the source of HCN. Indeed, for this model, best fits of $R_{\rm Obs}$ are obtained for parent scale lengths $<$ 100 km, implying production of HCN in the very inner coma. In contrast, when anisotropic outgassing is considered, the brightness distribution of HCN is consistent with a distributed source of HCN with a parent scale length of up to 1000 km (for pure extended HCN distribution) or even larger (composite case). However, $R_{\rm Obs}$ can also be explained with a near-nucleus production of HCN (Fig.~\ref{fig:HCN-rad}). The small dependency of $R_{\rm Obs}$ with $L_{\rm p}$ in the case of a composite origin of HCN is related to optical depth effects which are more important for small $L_{\rm p}$ values, producing an extended brightness distribution whereas the radial distritution is steep. In summary, the measured intensity ratio $R_{\rm Obs}$ by itself does not provide strong constraints on the parent scale length. However, the 
offset of the brightness peak is also sensitive to the HCN radial distribution. As shown in Fig.~\ref{fig:HCN-rad}, the observed large position offset of 3.5$''$ excludes a broad ($\psi$ $\simeq$ 20\tdeg) HCN jet in the composite case. If most of the HCN molecules were produced by an extended source, then both the observed peak offset and $R_{\rm Obs}$ can be explained providing $L_{\rm p}$ = 400--1000 km ($\psi$ = 30\tdeg), or $L_{\rm p}$ $<$ 150 km ($\psi$ = 10\tdeg). A large opening angle is best consistent with the distribution of the cloud of icy grains and CO$_2$ released from the illuminated part of the small lobe \citep {ahe+11}. If these grains are the source of HCN, then our model results suggest that the characteritic scale for production of HCN was typically 500--1000 km. Using the time scale of production of HCN determined in Sect.~\ref{sec-jet} of 1.7 h, we can determine the velocity of the icy grains producing HCN to be on the order of 100--200 m s$^{-1}$,  which is consistent with the determination of \citet{dra+12}, from a different approach. These values correspond to the velocity of 15--80 $\mu$m grains according to model (2) of Sect.~\ref{sec:dust}. \citet{Beer+06} computed the lifetime of pure and dirty icy grains made of dark material and 90\% of ice at a distance from the Sun of 1.09 AU, which is close to the distance of 103P/Hartley 2 on 4 November (1.06 AU). The time scale of 1.7 h is too short to invoke the production of HCN from pure icy grains. On the other hand, it corresponds to the lifetime of 600 $\mu$m dirty grains. Since the lifetime of the grains depends on their composition (but not significantly on the ice content) \citep{Gicquel+12,Beer+06} and that less absorbing material than the dark material considered by                \citet{Beer+06} are present in cometary grains (e.g., silicate), there might be no contradiction between the scale length and time scale of production of HCN from icy grains.

\begin{table*}
\begin{center}
\caption{Single-dish to interferometric flux ratios.}
\label{tab-ratios}
\begin{tabular}{l c c c c }
\noalign{\smallskip}
\hline
Line     & Date &   $F_{\rm SD}$$^a$    & $F_{\rm Int}$$^a$   &  $R_{\rm obs}$     \\ 
         &      &    Jy km s$^{-1}$ & Jy km s$^{-1}$ &   \\\hline
\noalign{\smallskip}
\hline
\noalign{\smallskip}
CH$_3$OH & 28 Oct. 2010$^b$&  6.34~$\pm$~0.35 & 0.507~$\pm$~0.065 & 12.5~$\pm$~2.1 \\
HCN $J$(1--0) & 23 Oct. 2010$^c$ &  0.68~$\pm$~0.09 & 0.042~$\pm$~0.020 & 16.2~$\pm$~7.9 \\
HCN $J$(1--0) & 4 Nov. 2010$^d$ & 0.84~$\pm$~0.13 & 0.044~$\pm$~0.015 & 19.1~$\pm$~7.0 \\ 
HCN $J$(3--2) & 5 Nov. 2010$^e$ & 7.49~$\pm$~0.80 & \phantom{0}0.24~$\pm$~0.050 & 31.2~$\pm$~8.1 \\
\hline
\noalign{\smallskip}
\end{tabular}
\end{center}

$^a$ Errors include absolute flux calibration uncertainties (10\% in On-Off and 5, 10, and 15\% in interferometric mode at 88.6, 157.2, and 265.9~GHz, respectively).
$^b$ Sum of the intensities of the six 157.2~GHz methanol lines. 
$^c$ Time range from 2 to 5 h UT in On-Off mode, and 1–-4 h UT in interferometric mode.
$^d$ Time range from 2 to 3.5 h UT in On-Off, and 1 --2.5 h UT, in interferometric mode.
$^e$ Whole data set acquired from 0 to 8.5 h UT.

\end{table*}

\section{Summary}
\label{sec-sum}

We observed comet 103P/Hartley 2 with the IRAM Plateau de Bure interferometer both in interferometric and single-dish modes. Observations were conducted on 23, 28 October, and on 4 and 5 November 2010, i.e., close to the time of the EPOXI flyby. The nucleus and dust thermal emissions were detected, as well as emission lines of HCN 
and CH$_3$OH. Taking advantage of complementary observations and published spin parameters and shape model, we interpret the observed time variability of gas production of 103P/Hartley 2, {and present two independent lines of evidence for production of HCN from grains.}   Our results can be summarized as follows:    

\begin{itemize}
\item Both the nucleus and the dust thermal emissions are found to contribute to the detected continuum emission at 88.6, 157.2, and 265.9 GHz. The amplitude and period of the sine-like thermal light curve observed on 23 October 2010 
show that the temporal variation of the continuum was driven by nucleus rotation. Depending on the date and frequency of the measurement, the nucleus thermal emission contribute from 30 to 55\% of the observed flux.
The dust thermal emission is analysed with a thermal model. Dust production rates are derived under various assumptions regarding dust composition, size distribution and grain velocities. We infer values in the range 830--2700 kg s$^{-1}$, which correspond to dust-to-gas ratios in the range 2--6. Large dust-to-gas ratios can be explained by the unusual activity of the comet given its size, which allows decimeter-sized particles or larger boulders to be entrained by subliming gases.

\item We measure temporal variations of the intensity of HCN and CH$_3$OH lines, which are consistent with 
measurements obtained by \citet{dra+12} with the IRAM 30-m telescope. In addition, interferometric maps obtained with a resolution of 100--500 km reveal a displacement of the HCN and CH$_3$OH peak emission with respect to the peak brightness of continuum emission. Whereas the daily-average displacement is South-East, motions along the  North-South axis are observed in time scales of a few hours. Time variations of the mean velocity of the line are also detected, though with less confidence than \citet{dra+12} because of moderate signal-to-noise ratios. We interpret these time variations induced by the rotation of the nucleus using geometrical considerations. The shape model of \citet{thomas+12} is used, and we consider the complex rotational state of 103P's nucleus, using the spin parameters and space orientation of the total angular momentum determined by \citet{belton+12}. We show that gas activity is modulated by the insolation of the small lobe of the nucleus. A delay of 
 1.7 h is observed between the HCN production curve and that of CO$_2$ measured by EPOXI \citep{besse+12}, suggesting that HCN is mainly produced by icy grains. The  similar phasing of the H$_2$O, HCN, and CH$_3$OH production curves shows that the three species have common sources. The three-cycle periodicity of the production curves can be explained by an overactivity of the small lobe in the longitude range 0--180$^{\circ}$. The time evolution of the velocity offset of the HCN lines, as well as the displacement of the HCN photocenter, are overall well reproduced. Measurements at times when the small lobe is in shadow suggest the presence of other localized sources of gas on the nucleus surface. The good correlation between the insolation of the small lobe and CO$_2$ production is consistent with CO$_2$ being produced from small depths below the surface.

\item Six methanol lines at $\sim$157.2~GHz are observed on the same date with the Plateau de Bure and the IRAM 30-m telescopes, allowing us to measure the CH$_3$OH rotational temperature with beam radii varying from 150~km and 1500~km.  The increase of temperature with increasing beam size (from 35$\pm$6~K to 46$\pm$2~K) can be explained by a change in the excitation conditions, with radiative processes becoming significant with respect to collisional excitation in the outer coma. The measured temperatures are lower than the values of 70--90~K  derived from infrared observations with smaller fields of view  \citep{del+11,mum+11,bonev+12}. A possible explanation is a rapid decrease of the gas kinetic temperature in the first tenths of kilometers above the surface as expected for an adiabatic expansion of the gas.

\item We investigate whether there is evidence for extended production of HCN in the Plateau de Bure data using as constraints the ratio between the interferometric and single-dish fluxes, and the position offset of the peak brightness in the maps. Our model results suggest that HCN molecules are mainly produced by a distributed source characterized by a scale length of typically 500--1000 km, implying a mean velocity for the icy grains producing HCN of 100--200 m s$^{-1}$. The time scale of production of HCN of 1.7 h is too short to invoke the production of HCN from pure icy grains.

\end{itemize}

 The present study complement results obtained from infrared spectroscopy which probed various molecules, including HCN and CH$_3$OH \citep{del+11,mum+11,kawa+13}. In most recorded infrared spectra, the gas emissions were more extended than dust, supporting production from icy grains. The spatial distribution of H$_2$O and CH$_3$OH were found similar to each other, but were generally different from C$_2$H$_6$, HCN and C$_2$H$_2$, indicating that there might be two distinct phases of ice in comet Hartley 2, one enriched in H$_2$O and CH$_3$OH, and another enriched in more volatile species (C$_2$H$_6$, C$_2$H$_2$, and HCN) \citep{del+11,mum+11,kawa+13}. Combining radio and infrared results, 
the production curves of these various molecules were approximately in phase, showing that the icy grains producing these molecules were principally originating from the small lobe of the nucleus. A large fraction of the water and methanol molecules were released from slowly sublimating pure icy grains \citep{fougere+13}. This scenario is consistent with the steady production rate ratios observed throughout the apparition \citep{mum+11,kawa+13}.

\section{Acknowledgements}
We thank M. Drahus, for providing us his data set, and M. Belton for useful discussions concerning the complex rotational state of 103P's nucleus. P. Thomas provided us the shape model and is greatly acknowledged. We also thank S. Besse for providing us unpublished information on H$_2$O and CO$_2$ light curves recorded by EPOXI, and M.R. Combi and N. Fougere who kindly communicated to us unpublished results of their 103P's model. This work is based on observations carried out with the Plateau de Bure Interferometer operated by the IRAM which is supported by INSU/CNRS (France), MPG (Germany), and IGN (Spain).  
CSO is supported by the NSF, award AST-0540882.
The research leading to these results  received funding from the European Community's Seventh Framework Programme (FP7/2007--2013) under grant agreement No. 229517.

\bibliographystyle{icarus}
\bibliography{bib103P.bib}

\end{document}